\documentclass[12pt]{article}
\pdfoutput=1

\usepackage[utf8]{inputenc}
\usepackage{cancel}
\usepackage{amssymb}
\usepackage{slashed}
\usepackage{soul}
\usepackage{tabularx}
\usepackage{xcolor}
\usepackage{booktabs}
\usepackage{subcaption} 
\usepackage{colortbl}
\usepackage{authblk}
\usepackage{graphicx}
\usepackage{diagbox}
\DeclareUnicodeCharacter{00A0}{ }

\newcolumntype{C}{>{\centering\arraybackslash}X}
\usepackage[T1]{fontenc}
\usepackage{epsfig}
\usepackage{latexsym}
\usepackage{graphicx}
\usepackage{amsmath}
\usepackage{amsfonts}   
\usepackage{amssymb}    
\usepackage{float}
\usepackage{bm}
\usepackage{url}
\usepackage{hyperref} 
\usepackage[nodisplayskipstretch]{setspace}
\def\lsim{\raise0.3ex\hbox{$\;<$\kern-0.75em\raise-1.1ex\hbox{$\sim\;$}}}
\def\gsim{\raise0.3ex\hbox{$\;>$\kern-0.75em\raise-1.1ex\hbox{$\sim\;$}}}

\def    \beq            {\begin{equation}}
\def    \eeq            {\end{equation}}
\def    \bea           {\begin{eqnarray}}
\def    \eea           {\end{eqnarray}}

\def\g2{{\rm GeV}^2}

\def\sw2{sin^2 \theta_w}

\def\a^tau{\alpha_{\tau}}

\def\beq{\begin{equation}}
\def\eeq{\end{equation}}
\def\beqa{\begin{eqnarray}}
\def\eeqa{\end{eqnarray}}

\newcommand{\newc}{\newcommand}
\newc\BR{BR}
\newc{\akappa}{A_{\kappa} }
\newc\deltagmtwo{\delta (g-2)_{\mu}} 
\newc\deltaamu{\Delta a_{\mu}}

\def\anti{\overline}

\newc{\haa}{BR\(h_1\to a_1 a_1\)}
\newc{\abb}{BR\(a_1\to b\anti{b}\)}
\newc{\hbb}{BR\(h_1\to b\anti{b}\)}
\newc{\abund}{\Omega h^2}
\newc\bsgamma{b\rightarrow s \gamma }
\newc\bxsgamma{\overline{B}\rightarrow X_{s}\gamma}
\newc\brbsgamma{\BR(\overline{B}\rightarrow X_s\gamma)}

\newc{\Fermi}{\textit{Fermi}-}

\allowdisplaybreaks

\graphicspath{{./Figures/Diagrams/}}
\usepackage{enumitem}

\setul{0.5ex}{0.3ex}
\definecolor{Blue}{rgb}{0,0.0,1}
\setulcolor{Blue}

\usepackage{array, makecell}
\usepackage{boldline}
\usepackage[nosort]{cite}

\usepackage{tikz}
\usepackage{mathtools}
\usepackage[capitalise]{cleveref}

\definecolor{lime}{HTML}{A6CE39}
\DeclareRobustCommand{\orcidicon}{%
    \begin{tikzpicture}
    \draw[lime, fill=lime] (0,0) 
    circle [radius=0.16] 
    node[white] {{\fontfamily{qag}\selectfont \tiny ID}};
    \draw[white, fill=white] (-0.0625,0.095) 
    circle [radius=0.007];
    \end{tikzpicture}
    \hspace{-2mm}
}
\newcommand{\orcid}[1]{\href{https://orcid.org/#1}{\orcidicon}}

\begin{document}

\begin{titlepage}

\thispagestyle{empty}

\def\thefootnote{\fnsymbol{footnote}}

\begin{flushright}
IFT-UAM/CSIC-24-89\\
\end{flushright}

\vspace*{1cm}

\begin{center}

\begin{Large}
\textbf{
Insights into Dark Matter Direct Detection Experiments: Decision Trees versus Deep Learning
}
\end{Large}

\vspace{1cm}

{\sc
Daniel~E.~L\'opez-Fogliani$^{a, b}$\orcid{0000-0001-8564-6676}%
\footnote{{\tt \href{mailto:daniel.lopez@df.uba.ar}{daniel.lopez@df.uba.ar}}}%
, Andres D. Perez$^{c, d}$\orcid{0000-0002-9391-6047}%
\footnote{\tt \href{mailto:andresd.perez@uam.es}{andresd.perez@uam.es}}%
, Roberto~Ruiz~de~Austri$^{e}$\orcid{0000-0003-3688-9609}%
\footnote{{\tt \href{mailto:rruiz@ific.uv.es}{rruiz@ific.uv.es}}}%
}

\vspace{0.5truecm}

{\sl
$^a$Instituto de F\' \i sica de Buenos Aires, UBA \& CONICET, Departamento de F\' \i sica, Facultad de Ciencia Exactas y Naturales, Universidad de Buenos Aires, 
 1428 Buenos Aires, Argentina

\vspace*{0.15cm}

$^b$Pontificia Universidad Cat\'olica Argentina, Av. Alicia Moreau de Justo 1500, 
1107~Buenos~Aires, Argentina

\vspace*{0.15cm}

$^c$Instituto de F\' \i sica Te\'orica, IFT-UAM/CSIC, Campus de Cantoblanco, 28049 Madrid, Spain

\vspace*{0.15cm}

$^d$Departamento de F\' \i sica Te\'orica, Universidad Aut\'onoma de Madrid, Campus de Cantoblanco, 28049 Madrid, Spain

\vspace*{0.15cm}

$^e$Instituto de F\' \i sica Corpuscular CSIC-UV, c/Catedr\'atico Jos\'e Beltr\'an 2, 46980 Paterna (Valencia), Spain
}

\vspace*{2mm}

\end{center}

\vspace{0.1cm}

\renewcommand*{\thefootnote}{\arabic{footnote}}
\setcounter{footnote}{0}

\begin{abstract}
\noindent 

The detection of Dark Matter (DM) remains a significant challenge in particle physics. This study exploits advanced machine learning models to improve detection capabilities of liquid xenon time projection chamber experiments, utilizing state-of-the-art transformers alongside traditional methods like Multilayer Perceptrons and Convolutional Neural Networks. We evaluate various data representations and find that simplified feature representations, particularly corrected S1 and S2 signals as well as a few shape-related features including the time difference between signals, retain critical information for classification. Our results show that while transformers offer promising performance, simpler models like XGBoost can achieve comparable results with optimal data representations. We also derive exclusion limits in the cross-section versus DM mass parameter space, showing minimal differences between XGBoost and the best performing deep learning models. The comparative analysis of different machine learning approaches provides a valuable reference for future experiments by guiding the choice of models and data representations to maximize detection capabilities.

\end{abstract}


\end{titlepage}

\tableofcontents


\section{Introduction}

The detection of DM beyond gravitational interaction is one of the most important experimental challenges in particle and astroparticle physics. From the theoretical side, to elucidate DM composition is one of the main motivations for physics beyond the Standard Model of Particle Physics. In this context, a Weakly Interacting Massive Particle, WIMP, is one of the best motivated DM candidates~\cite{Roszkowski:2017nbc,Schumann:2019eaa}.

Direct Detection experiments search for WIMP dark matter particles scattering with Standard Model (SM) particles through the measurement of the transferred energy inside a detector~\cite{Billard:2021uyg,Cebrian:2022brv}. Several current and near future collaboration employ ton-scale detectors with different target materials such as xenon (XENON~\cite{XENON:2020kmp,XENON:2023cxc}, LZ~\cite{LZ:2018qzl,LZ:2022lsv}, and PandaX~\cite{PandaX:2018wtu,PandaX-4T:2021bab}, and the proyected DARWIN~\cite{DARWIN:2016hyl}) or argon (DEAP-3600~\cite{DEAP:2019yzn} and the planned DarkSide-20k~\cite{Manthos:2023swh,Zani:2024ybb} and Argo~\cite{DarkSide20k:2020ymr}). In addition, several experiment currently taking data, employ sodium iodine (COSINE~\cite{COSINE-100:2021zqh,COSINE-100:2022dvc} and ANAIS~\cite{Coarasa:2024xec}, and the future COSINUS~\cite{Angloher:2016ooq,COSINUS:2023kqd}, SABRE~\cite{SABRE:2018lfp,SABRE:2022twu,Mariani:2022yat}, and PICOLON~\cite{PICO-LON:2015rtu,Fushimi:2024lpn}) to test the DAMA annual modulation signal~\cite{BERNABEI2020103810}.

To distinguish a WIMP signal from the background is crucial, and new machine learning (ML) techniques could play a essential role in the near future. In fact, ML is showing promising results (see e.g.~\cite{Guest:2018yhq,Albertsson:2018maf,Carleo:2019ptp,Bourilkov:2019yoi,Feickert:2021ajf,Schwartz:2021ftp,Arganda:2021azw,Karagiorgi:2021ngt}) in the field of particle physics and astroparticle physics. 
In particular, in DM direct detection phenomenology both supervised and semi-supervised approaches for direct detection were discussed in \cite{Khosa:2019qgp, Herrero-Garcia:2021goa, Bras:2022ejk, Cerdeno:2024uqt}. Several collaborations, like LUX \cite{LUX:2022vee} and CRESST \cite{CRESST:2022qor}, also have applied machine learning techniques to their data analysis. In \cite{Coarasa:2021fpv,Coarasa:2022zak} was suggested to improve ANAIS-112 sensitivity to DAMA/LIBRA signal using machine learning techniques. In the case of directional dark matter experiments, DRIFT collaboration release \cite{DRIFT:2021uus}, where the improvement of experiment sensitivity by the use of machine learning techniques was discussed. Additionally, the application of deep learning to distinguish between DM signals vs several backgrounds in the context of the proposed NEWSdm directional experiment was discussed in \cite{Golovatiuk:2020krw}.

Naturally, detector simulators including accurate background and signal description are mandatory in order to make possible the analysis at the detector level. In this work we focus on the XENON experiment \cite{XENON:2019izt}, taking advantage of the public simulator developed by the collaboration \cite{PaX,laidbax,blueice}, although in principle our conclusions can be applied to any liquid xenon time projection chamber experiment.

For future applications and developments in the field, it is crucial to discriminate the efficiency of different classifier algorithms. We focus our attention on this point in this work, utilizing state-of-the-art machine learning techniques like transformers, alongside traditional methods like Boosted Decision Trees, Deep Neural Networks and Convolutional Neural Networks. Additionally, we compare the performance of those classifiers applying various data representations to describe the detector response. While transformers offer a promising performance, simpler and much faster models like XGBoost achieve comparable results in most cases. We find that the simplified feature data representations, besides offering an easier interpretation, retain critical information for classification and thus provide the best performances. Finally, we use the ML classifier outputs to derive exclusion limits in the cross-section versus DM mass parameter space, showing minimal differences between XGBoost and the best performing deep learning models.

This paper is organized as follows. In Section~\ref{sec:DMDD} we briefly describe the used experiments, proposed WIMP events and the types of backgrounds. Section~\ref{sec:Data} is devoted to the process of simulating events. We also discuss the different data representations that we compare. The architectures used, the training and testing procedure, are explained in Section \ref{sec:Training}. In Section~\ref{sec:results}, we discuss the results. Finally, in Section \ref{sec:conclusions} we give the conclusion of this work.

\section{Dark Matter Direct Detection}
\label{sec:DMDD}

Liquid xenon time projection chambers (TPCs)~\cite{Aalbers:2022dzr} are one of the leading technologies in the search for finding and identifying the nature of DM. Collaborations like XENON~\cite{XENON:2020kmp,XENON:2023cxc}, LZ~\cite{LZ:2018qzl,LZ:2022lsv}, and PandaX~\cite{PandaX:2018wtu,PandaX-4T:2021bab} employ TPCs in direct detection experiments searching for DM particles scattering with the detector material to measure the transferred energy. They are designed to be sensitive to WIMP masses $O(\text{GeV-TeV})$ for which they establish the world leading constraint, but the measurements can also be applied to constraint non-vanilla DM scenarios~\cite{XENON:2022ltv,LZ:2023poo,PandaX:2022xqx}.

Generally, a liquid xenon TPC consists of a central liquid xenon volume surrounded by light reflectors. Two arrays of photomultiplier tubes (PMTs) are located at the top and bottom of the TPC to collect the light signals. A single interaction coming from an incident particle would deposit energy in the liquid xenon, producing two signals: a prompt scintillation light and delayed ionization electrons. The scintillation light is collected by the PMT arrays immediately, due to the reflecting material on the walls, and is denominated signal S1. However, the ionization electrons have to be extracted from the liquid xenon. To do so, a cathode and a gate on the bottom and top parts of the liquid xenon, respectively, generate an electric field. When the drifting electrons pass through the gate, they are extracted with another electric field into a gas xenon phase of the experiment, located just above the liquid xenon region. The extracted electrons produce a light signature that is detected by the PMTs, and is called S2 signal. The position of the interaction in the horizontal plane is determined by the S2 top-array PMT hit pattern, while the time difference between the S1 and S2 signals define the depth of the event. The energy deposited is reconstructed using S1 and the S2 bottom-array PMT. Importantly, the S1 signal has to be corrected (cS1) for position-dependent light collection efficiency due to geometric effects, as well as the S2 signal (cS2) taking into account the electron lifetime and position-dependent inhomogeneous electroluminescence amplification.

The interaction sources can be classified into two types: electronic recoils (ER) when the incoming particle scatters off the atomic electrons, and nuclear recoils (NR) when the scatter is produced off the xenon nuclei. There are many contributions to the ER background, the main one comes from radioactive elements, $\beta$ decays in the decay chains of Radon, Krypton and Xenon isotopes that are homogeneously distributed in the active detector volume. Additional sources include radioactivity in the detector materials (also called surface events), and solar neutrinos scattering of electrons. The NR background consists on neutrons originated in spontaneous fission reactions and on coherent elastic neutrino-nucleus scattering (CE$\nu$NS) mainly from the $^8B$ channel from the Sun. Finally, another contribution comes from accidental coincidences, the pairing of S1-S2 signals from different sources that are identified as originating from a single event.

Regarding the signal, throughout this work we are going to consider DM with spin-independent cross-section within the standard WIMP paradigm, for which we expect a NR type interaction.

\section{Data Generation}
\label{sec:Data}

In this work, to study the different algorithms to discriminate between signal and background, we considered the characteristics of XENON1T~\cite{XENON:2019izt}. To create a dataset of events, we used a simulator developed by the XENON Collaboration~\cite{PaX,laidbax,blueice}. The package includes templates for all the backgrounds described in the previous section. Although these are not employed by the XENON Collaboration in their published analysis, we have checked that the templates and the simulator reproduce the available results~\cite{XENON:2017vdw,XENON:2018voc} regarding the expected number of events of each source. We generated a set of $30$k background events, taking into account the relative weight of each source.

To produce signal events, we need to provide the simulator with the differential rate of the particular WIMP candidate that we want to test, which we computed using \texttt{wimprates}~\cite{jelle_aalbers_2022_7041453} considering a xenon detector in the standard DM halo model. We generated several datasets of $30$k events for different WIMP mass, $m_{dm}=[10, 20, 35, 50, 100, 200, 500]$ GeV, but fixing the WIMP-nucleon cross-section to $\sigma^{SI}_{dm-\mathcal{N}}=10^{-45}$ cm$^2$. This allows to ease the computation time which is possible since the differential rate, and therefore the number of expected events, depends on linearly with the cross-section (see Eq.~\ref{eq:diff_rate} in the Appendix~\ref{Appen:signal_rates}). To account for different cross-sections, we can reweigh the expected number of signal events.

In Table~\ref{tab:numberofevents} we show the expected number of background and signal events computed with the simulator for a XENON1T experiment with 1 ton yr exposure, considering 350 days of data taking and a fiducial mass of 1042 kg liquid xenon contained in a cylindrical volume of 39.85 cm radius and 70 cm height in the center of the TPC enclosure. As an example, the WIMP has $m_{dm}=500$ GeV, $\sigma^{SI}_{dm-\mathcal{N}}=10^{-45}$ cm$^2$.

\begin{table}[t]
    \centering
    \begin{tabular}{c|c}
  \hline
        Source & Expected number of events \\
    \hline
        ER & 634.55 \\
        Surface events & 5.34 \\
        Radiogenic neutrons & 0.60 \\
        CE$\nu$NS & 0.01 \\
        Accidental coincidences & 2.26 \\
        WIMP & 33.25 \\
        \hline
    \end{tabular}
    \caption{Number of expected events for each source in a XENON1T experiment with 1 ton yr exposure, computed with the simulator. For the WIMP model we considered $m_{dm}=500$ GeV, $\sigma^{SI}_{dm-\mathcal{N}}=10^{-45}$ cm$^2$ as a benchmark.}
    \label{tab:numberofevents}
\end{table}

We would like to highlight that we take XENON1T as a benchmark because there is a public software that describes the interactions. This guarantees the reproducibility of the procedures described in this work, since it is easy to check that the simulator reproduces the published results without significant changes to the code. Moreover, we are interested in a general study useful for any direct detection experiment based on liquid xenon dual-phase TPCs. Taking into account that the working principle is the same, the results obtained with this simulator regarding the performance of ML classifiers and the comparison between different data representations can be applied for current and next-generation liquid xenon experiments like DARWIN~\cite{DARWIN:2016hyl} or PandaX-xT~\cite{PandaX:2024oxq}. Therefore, we expect that a more accurate description of the backgrounds modifying the templates or the use of another simulator with the particular specifications of a different apparatus will not change significantly our conclusions.

\subsection{Data Representation}
\label{sec:DataRepresentation}

The events generated with the simulator were stored in different data representations, we analyse and compare them in Section \ref{sec:results}. As discussed below, we divide the representations in two groups: raw data and processed data.

\subsubsection*{Raw data: images (dataset A)}

\begin{figure}
  \centering
  \includegraphics[width=0.4\textwidth]{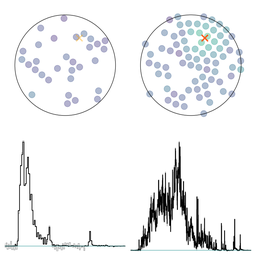} \hspace{0.75cm}
  \includegraphics[width=0.4\textwidth]{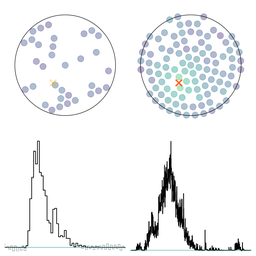}
\caption{Dataset A1 examples (raw data as images). The left image corresponds to a signal event with $m_{dm}=500$ GeV, $\sigma^{SI}_{dm-\mathcal{N}}=10^{-45}$ cm$^2$, and the right image to an ER background event. In each image, the upper panels represent the hit patterns on the bottom and top PMTs, left and right respectively, and the lower panels show the time pulse of the largest S1 (left) and S2 (right) peaks.}
\label{fig:images}
\end{figure}

Within the first group, we consider the response of the detectors as images, as analysed in \cite{Khosa:2019qgp, Herrero-Garcia:2021goa}. We call this representation dataset A1. In Fig.~\ref{fig:images} we show two examples, the left image corresponds to a signal event with $m_{dm}=500$ GeV, $\sigma^{SI}_{dm-\mathcal{N}}=10^{-45}$ cm$^2$, and the right image to an ER background event. Notice that both images are very similar (we have removed the axis labels to ease image classification, but signal and background events have also similar magnitudes). Each event is characterized by an image composed of four panels, the first two correspond to the responses or hit patterns of the bottom (upper left panel) and top (upper right panel) PMTs, where the relative position of the PMTs is shown by a circle and the intensity of the signal detected by each PMT is represented by a colour gradient. Additionally, the lower panels correspond to the largest S1 (left) and S2 (right) intensity peaks as a function of time, where the intensity of the signal corresponds to the total amount of light collected by all the PMTs. The main advantage of this representation is to make explicit the correlation between the responses of each PMT that could be exploited by a ML algorithm like a CNN. On the other hand, one of the main drawbacks of working with images is that we need to generate a large dataset, which can be computationally very expensive to analyze and to store. We compressed the images generated with the simulator and tested different resolutions $75\times75$, $128\times128$, and $256\times256$ pixels, and found no significant differences between them, therefore we present the results using $75\times75$ pixels images.

\subsubsection*{Raw data: arrays (dataset B)}

The second type of raw data that we consider also includes the same four detector responses, but instead of working with images, we store four vectors per event and call this representation dataset B1. The first vector corresponds to the intensity of the bottom hit pattern, i.e. a number in the interval $[0,1]$ representing the intensity of each of the 121 PMTs, the second one with the top hit pattern, a vector with the intensity of the 127 PMTs, the third vector with the intensity of the largest S1 peak normalized to 1 as a function of time, and the fourth to the largest S2 peak. In this representation the third and fourth vectors have different length for every event because the program takes a measurement of the intensity (sample time) every 10 ns while the time duration of each signal can vary. To be able to analyse this information with ML we add extra zero values, a method called zero padding, to obtain vectors of the same length (we apply this procedure to the S1 and S2 peaks separately).

\subsubsection*{Processed data: S1 and S2 peak reduced features (dataset C)}

Regarding the processed data, we construct a set of reduced features that are relevant shape-related quantities describing the S1 and S2 peaks, such as integrated areas and lengths. Some of the features considered in this work are based on the ones taken in Ref.~\cite{Bras:2022ejk}. The set of shape-related quantities, denoted dataset C1, has the following 14 quantities
\begin{itemize}
    \item pA: total integrated area from the start to the end of the S1 (S2) pulse,
    \item pH: maximum amplitude of the S1 (S2) pulse,
    \item pHT: time at which the S1 (S2) pulse reaches its maximum amplitude with respect to the initial time of the pulse,
    \item pL: time difference between start and end of the S1 (S2) pulse,
    \item pL90: length time with 90\% of the S1 (S2) pulse area, from 5\% to 95\% integrated area,
    \item pRMSW: S1 (S2) pulse root-mean-square (RMS) width,
    \item pHTL: ratio between the time at which the S1 (S2) pulse reaches its maximum amplitude and the time difference between start and end of the S1 (S2) pulse.
\end{itemize}
We also construct another set, called dataset C2, by including an additional feature,
\begin{itemize}
    \item $\Delta t$: time difference between S1 and S2 pulses.
\end{itemize}
Notice that the information about the time separation between pulses is not available in either of the previous raw data representation by construction. Therefore, we will combine $\Delta t$ with both raw data representations, images and arrays, to be able to compare datasets fairly, and denote the respective datasets A2 and B2.

\begin{figure}
  \centering
  \includegraphics[width=1\textwidth]{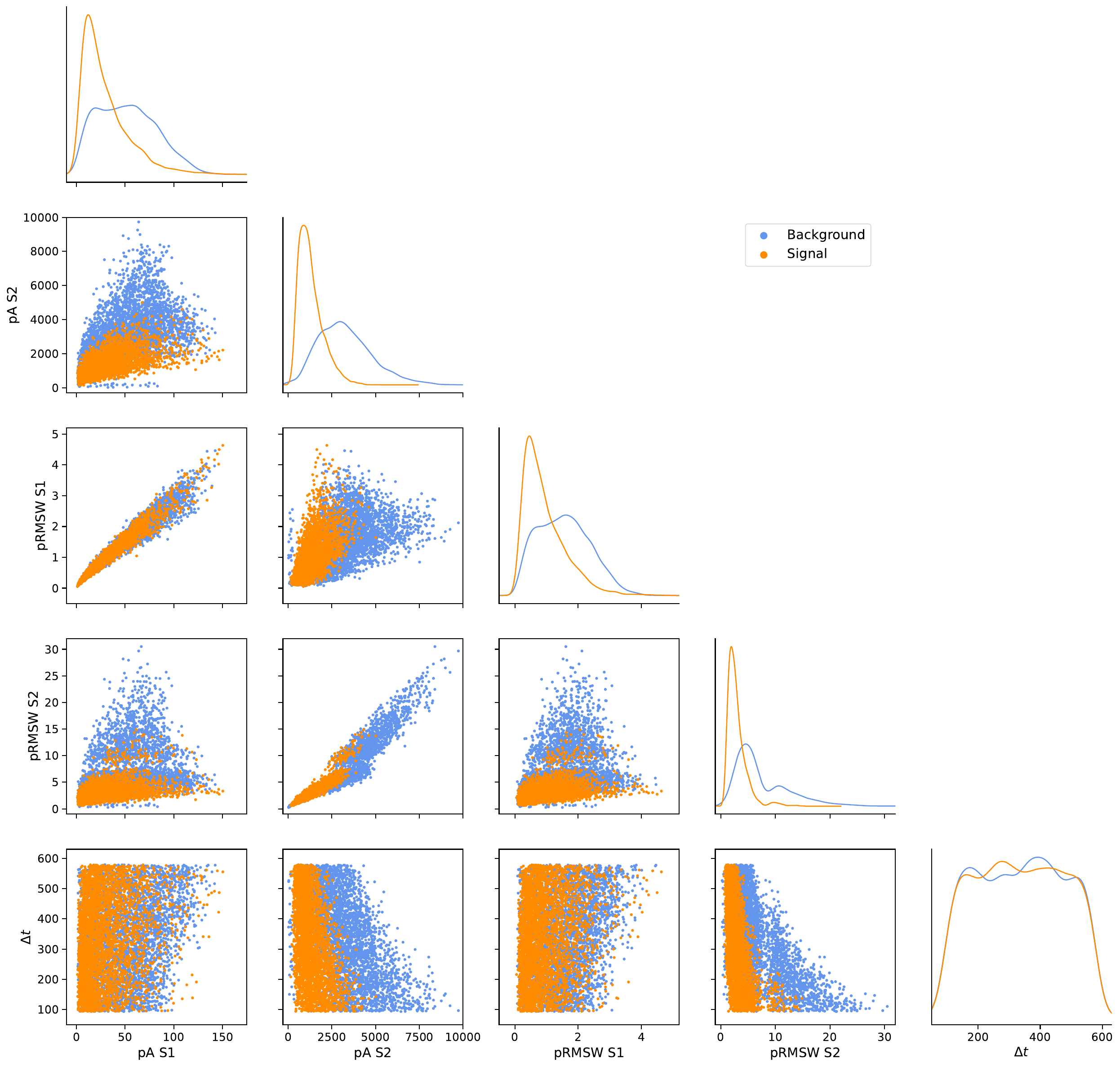} 
\caption{Correlations between the dataset C3 variables (five most important reduced features for classification according to XGBoost ordering), considering a signal event with $m_{dm}=500$ GeV.}
\label{fig:correlations}
\end{figure}

Furthermore, we consider a subset of the resumed features, called dataset C3, that corresponds to the five most important features determined by the XGBoost classifier for the discrimination between background and signal with $m_{dm}=500$ GeV. As we will show in Section~\ref{sec:results}, this subset retains enough information to provide an excellent performance. In Fig.~\ref{fig:correlations} we show the distributions and correlations (which are independent of the classifier) of the features in dataset C3, using the same number of background and signal events.

\subsubsection*{Processed data: cS1 and cS2 (dataset D)}

\begin{figure}
  \centering
  \includegraphics[width=0.5\textwidth]{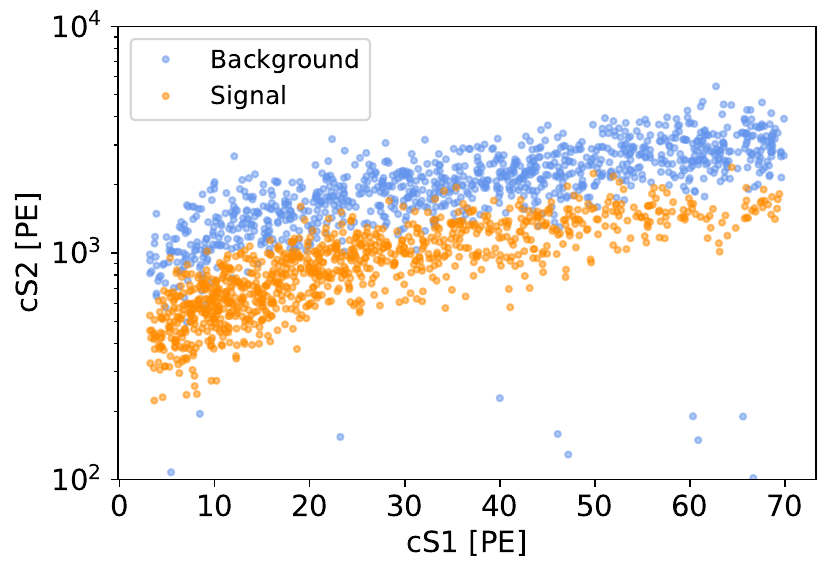} 
\caption{1k signal and 1k background events in the cS1-cS2 plane (dataset D). As an example, we show signal events with $m_{dm}=500$ GeV.}
\label{fig:cs1cs2}
\end{figure}

Finally, we also consider cS1 and cS2 and denote this representation as dataset D. As explained in Section~\ref{sec:DMDD}, these high-level features are similar to the total integrated area of S1 and S2 pulses (also to the sum of the intensity of all the PMTs) but corrected by factors that take into account the geometric effects of the TPC, position-dependent inhomogeneous effects, light collection efficiency, the drift electron lifetime under the drift field. For more details about the modelling of these corrections, see Ref.~\cite{XENON:2019izt}. In this work, we use the default values included in Ref.~\cite{PaX,laidbax,blueice}. The information in the cS1-cS2 plane is the one used by the collaboration using liquid xenon experiments to set limits to the DM parameter space, and naturally require a huge effort to characterize the detector response and to model the correction factors. The main advantage of this representation is its simplicity that allows a relative fast analysis even with complex models and ease the interpretation of the results. One of the goals of this work is to compare this simple representation under the scrutiny of state-of-the-art ML algorithms. Comparing the discrimination power of the resumed features and the raw data representations, we can check if the cS1-cS2 plane retains all the useful information. Moreover, this comparison also can be applied by the experimental collaborations to cross check the modelling of the S1 and S2 correction factors. In Fig.~\ref{fig:cs1cs2} we show an example with 1k signal and 1k background events in the cS1-cS2 plane, considering $m_{dm}=500$ GeV.

\section{Architecture and Training}
\label{sec:Training}

\subsubsection*{Boosted Decision Tree: XGBoost}

A gradient boosting decision tree (GBDT) algorithm employs, in addition to the boosting, gradient descent to minimize a differentiable loss function. XGBoost, or eXtreme Gradient Boost, is an optimized distributed gradient boosting algorithm that utilizes the second-order derivatives of the convex loss function, in contrast to a general GBDT algorithm where only the first-order derivative is used.

We divide the data into three subsets: 60$\%$ of the data is used for training, 15$\%$ for validation and the remaining 25$\%$ for testing. To avoid overfitting, we used an early stopping of 50 iterations that ends the training if the performance of the classifier does not improve when tested over the validation dataset. We also performed a grid search to find the optimal set of hyperparameters that maximizes the AUC. We have found that in most cases the default parameters produces results very close to the ones obtained with the optimal hyperparameters, that usually involves a maximum depth of a tree equal to 5, learning rate of 0.01, 2500 estimators, and a L2 regularization term on weights of 0.01.

\subsubsection*{Deep Neural Network}

Deep Neural Network (DNN) is a feed-forward multi-layered structure in which each layer is composed of multiple nodes, and can model complex non-linear functions. In this work we used fully connected layers, i.e. each node takes as input the output (involving a non-linear function called activation function) of all the nodes of the previous layer. For our classification task, the final layer (output layer) consist of a single node that ideally should reproduce the label of the data feed to the neural network. Then, during training, the weights of each node is updated by a procedure called back propagation.

We divide the data into three subsets: 60$\%$ of the data is used for training, 15$\%$ for validation and the remaining 25$\%$ for testing. We use the Adam optimizer to train the models with the default parameters, an early stopping of 50 iterations, and a batch size of 128. Additionally, a dropout of 0.2 was set after each hidden layer. We have not found significant differences varying those hyperparameters around the mentioned values. We have tested several activation functions and decided to employ Leaky-ReLU with alpha of 0.05 in most cases since it produces stable results in a relative short period of time. Finally, to optimize the AUC we varied the number of hidden layers from 1 to 8 using a different number of nodes (powers of 2, up to 512) in each layer.

\subsubsection*{Convolutional Neural Network}

Convolutional Neural Network (CNN) is a type of feed-forward multi-layered structure often used on images since adjacent pixel information can be exploited via filter, or kernel, optimization. Unlike DNNs, the hidden layers are not only fully connected layers, but a set of several convolutional layers performing a dot product of the convolution kernel with the layer's input matrix. Since the output of each convolutional layer is a set of images, they are followed by other layers such as pooling layers to reduce the dimensions of the data by combining the output of a few nodes into a single value. Finally, the output is flattened and a set of fully connected layers in a DNN-like structure is used to provide a prediction.

Each kernel is a matrix that scans the image. When a specific feature is found that corresponds to the kernel, the output of the convolutional layer is high. Then, during the training, the CNN learns different kernel weights such that each one can recognize a different feature relevant to the classification.

As in the DNN case, we divided the data into three subsets with the mentioned proportions. We used the same optimizer, early stopping, batch size, dropout values, and the same Leaky-ReLU activation function. However, the number of convolutional layers (1 to 5), number (4,8,16, or 32) and size of kernels (2$\times$2, 3$\times$3, or 5$\times$5), max-pooling size (2$\times$2 or 3$\times$3), and number (1 to 5) and size (2 to 32 nodes) of the fully connected layers, were optimized to maximize the AUC in each particular case and data representation.

\subsubsection*{Transformer models}

\begin{itemize}

\item Vision Transformer (ViT): For the image representation of raw data, we use a slightly modified version of the ViT model \cite{beyer2022better}. Modifications include the use of global average pooling instead of a class token and fixed 2D sin-cos position embeddings, among other adjustments to optimize performance for our particular application.

\item FormerTime Model: For the array representation of raw data, we use the FormerTime model \cite{cheng2023formertime}, which integrates both convolutional networks and transformers. This hybrid approach overcomes the limitations of modelling long-range dependencies and fixed-scale representations. The model features a hierarchical network architecture and a customized transformer encoder with a temporally reduced attention layer and contextual position coding, which increases classification, performance and efficiency.

\item TabTransformer: For summarized features, we use the TabTransformer model \cite{huang2020tabtransformer}, which is tailored for supervised and semi-supervised learning on tabular data. This model utilizes self-attention based transformers to convert categorical feature embeddings into robust contextual embeddings to improve prediction accuracy. The architecture includes a column embedding layer, multiple transformation layers and a multi-layer perceptron for the final predictions.

\end{itemize}

For the transformer models, we divide the data into three subsets: 60$\%$ of the data is used for training, 15$\%$ for validation and the remaining 25$\%$ for testing.
We use the Adam optimizer to train the models. The initial learning rate is set to $10^{-4}$, and to further refine the learning rate during training, we use a learning rate planner based on the plateau of the validation loss. If the validation loss does not improve for a certain number of epochs, the learning rate is reduced by a factor of two to allow for finer adjustments and to avoid possible local minima. Finally, when the learning rate reaches a value of $10^{-7}$, the training is aborted. In addition, we also performed a grid search to find the optimal set of hyperparameters that optimize the AUC.

\subsubsection*{Input shapes and incorporation of $\Delta t$}

To analyse the images with XGBoost and DNN algorithms, we have reshaped each dataset by flattening the original 2-dimensional matrix into a 1-dimensional array. On the other hand, to use CNN methods with 1-dimensional datasets (e.g. the reduced features), we employed 1-dimensional convolutional layers.

Finally, to incorporate the information about the time difference between the S1 and S2 pulses, $\Delta t$, into any 1-dimensional dataset, we simply extended the array in each event by including the corresponding $\Delta t$ value. Then, the shape of the input algorithm increases by~1. In the case of the CNN and ViT methods, whose inputs are 2-dimensional images, we incorporate the information about $\Delta t$ in the first fully connected layer of the ML algorithm (after the last convolutional layer when the output is flattened) by increasing the input of the fully connected layer by 1.\\

For further details about the training of the ML classifiers see Appendix~\ref{Appen:MLtraining}. All the code used in this work is available at~\cite{DDcode}.

\section{Results}
\label{sec:results}

\subsection{Classifiers performance}

\begin{table}[t]
    \centering
    \begin{tabular}{c|c|c||c|c|c|c}
  \hline
        \multicolumn{3}{c||}{} & XGBoost & DNN & CNN &  Transformer \\
  \hline
  \hline
  
        \textbf{A1} & Raw data & only raw & 0.919 & 0.924 & 0.918 &  \bf{0.930}$^*$ \\
  \cline{1-1}\cline{3-7}
        \textbf{A2} & (images) & + $\Delta t$ & 0.942 & 0.958 & 0.951 & \bf{0.965}$^*$ \\

  \hline
        \textbf{B1} & Raw data & only raw & 0.954 & 0.960 & 0.953 & \bf{0.963}$^\dagger$ \\
  \cline{1-1}\cline{3-7}
        \textbf{B2} & (arrays) & + $\Delta t$ & 0.969 & 0.978 & 0.968 &  \bf{0.980}$^\dagger$ \\

  \hline
        \textbf{C1} & Resumed & 14 features & 0.956 & 0.956 & 0.955 & \bf{0.966}$^\ddagger$  \\
  \cline{1-1}\cline{3-7}
        \textbf{C2} & features & + $\Delta t$ & 0.980 & 0.980 & 0.979 & \bf{0.982}$^\ddagger$  \\
  \cline{1-1}\cline{3-7}
        \textbf{C3} & & 4 main + $\Delta t$ & 0.980 & 0.979 & 0.972 & \bf{0.982}$^\ddagger$  \\

  \hline
        \textbf{D} & \multicolumn{2}{c||}{cS1, cS2} & 0.980 & 0.980 & 0.980 & 0.980$^\ddagger$  \\

  \hline
  \hline
  \multicolumn{3}{c||}{Uncertainty} & $\pm$ 0.001 & $\pm$ 0.002 & $\pm$ 0.003 & $\pm$ 0.002  \\
  \hline
    \end{tabular}
    \caption{AUC value for different classifier algorithms and data representations. For the signal, we considered a WIMP with $m_{dm}=500$ GeV as benchmark. The * denotes results obtained with the ViT model, $\dagger$ with the FormerTime model and $\ddagger$ with the tabTransformer model. }
    \label{tab:res}
\end{table}

In Table~\ref{tab:res} we present one of the main results of our manuscript comparing the AUC, area under the receiver operating characteristic (ROC) curve, of different ML classifiers and data representations. The AUC $\in [0.5,1]$ is a common quantity to evaluate the performance of a binary classifier, where AUC=1 indicates a perfect classification and AUC=0.5 a completely random result. To determine the AUC value and uncertainty, we have trained and tested each classifier 5 times using resampling with replacements, and report the mean and one standard deviation as final values. We found that the uncertainty is mainly associated with the ML method, therefore in Table~\ref{tab:res} we show one error value per column.

We have trained and tested the algorithms using a WIMP with $m_{dm}=500$ GeV as benchmark, but we have checked that a similar trend is obtained for different WIMP masses. Additionally, the WIMP cross-section is not relevant to determine the classification performance. This is because we are analysing and trying to distinguish the detector response in an event-by-event basis, and the cross-sections is related to the total expected number of events, but not to the transferred energy in a single elastic scattering nor to the shape of its distribution, as shown in Appendix~\ref{Appen:signal_rates}.

We can see in Table~\ref{tab:res} that the worst results correspond to the dataset A1 (first row) with AUC$\sim 0.92$. These results correspond to images with $75\times75$ to ease the computation, but with $128\times128$, and $256\times256$ pixels we found no significant differences. The main advantage of the images is to keep the spatial correlations, specially in the hit patterns of the top and bottom PMTs (see in Fig.~\ref{fig:images} the non-trivial distribution of the PMTs). However, the possible benefit is out-weighted by the increased complexity of analysing an image, which implies an exponentially larger number of parameters to be tuned by the ML classifiers. Considering the two algorithms that are designed to take advantage of such spatial correlations, the CNN and transformer, the first one gives the lowest performance of all the method reported, while the more complex transformer is able to barely surpass the rest. 

Comparing the results using both raw data representations, dataset A1 (images) versus dataset B1 (arrays), we notice an overall increase of the classifier performance when using the array representation. Recall that in the latter, one event is described with 4 arrays, two containing the response of each PMT (bottom and top) and two with the time series of the main pulses, i.e. the same information as in the image representation, but without the spatial correlation of the PMT. As explained before, this simpler representation not only allows for a significantly faster training with fewer parameters to tune but is more adequate, showing that the PMT spatial correlations do not provide significant information for the classification task. As before, the state-of-the-art transformer is the best performing algorithm, followed by the fully connected DNNs.

We want to highlight that we have also studied different combinations between the raw data representations. Instead of considering a single image composed of four panels, like the dataset A1 examples shown in Fig.~\ref{fig:images}, we have analysed each panel independently, and also used them to train 4 different ML classifiers that we combine afterwards into a single fully connected layer to provide a single classification. Moreover, we replaced some panels with the array representation, mixing datasets A1 and B1. For example, we used the images of the PMT responses to keep the spatial correlations combined with the arrays of the time series. To ease the interpretation of our results we do not show these possibilities, since there are lots of combinations and algorithm configurations, but no significant gain was achieved.

Regarding the processed data representations, we can see that the 14 shape-related resumed features (dataset C1) that describe the S1 and S2 pulses present a similar or slightly higher performance than the raw data as arrays (dataset B1), indicating that these set of variables capture the signal characteristics properly. Naturally, the smaller set of features implies smaller ML algorithms and faster training times.

\begin{figure}
  \centering
  \includegraphics[width=0.55\textwidth]{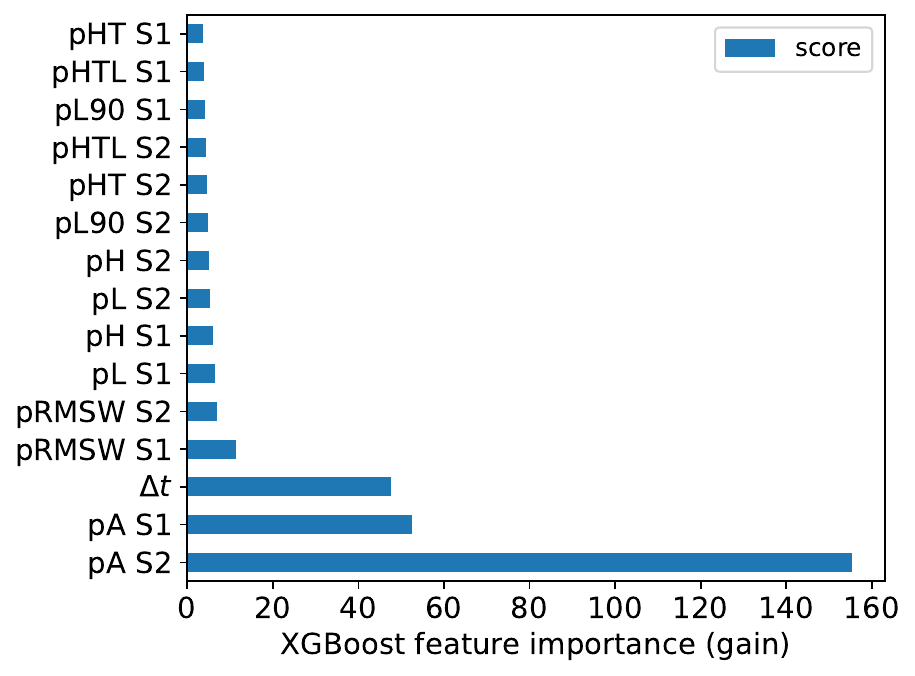}
\caption{Feature importance considering dataset C2 (14 pulse shape-related resumed features and $\Delta t$) for XGBoost classifier. WIMP with $m_{dm}=500$ GeV as benchmark.}
\label{fig:featureimportance}
\end{figure}

One feature that we have not discussed so far is the inclusion of the time difference between the S1 and S2 signals, $\Delta t$. We notice a significant increase in performance when comparing datasets without and with this feature: A1 versus A2, B1 versus B2, and C1 versus C2, respectively.

If we focus on the shape-related quantities, looking only at the bottom right panel of Fig.~\ref{fig:correlations}, one might assume that the addition of $\Delta t$ provides no extra discrimination power and should not be included in the training procedure, as both its background and signal distributions are similar. However, the impact of $\Delta t$ becomes evident when examining its correlations with other features, especially those related to S2 (the signal originating from the ionization electrons drifting towards the upper part of the detector). For instance, in the second panel of the last row of Fig.~\ref{fig:correlations}, for pA S2 $\sim 2500$ PE, events with $\Delta t \gtrsim 350$ $\mu$s are more likely to be background. In contrast, when looking solely at the distributions of the pA S2 feature (second panel of the second row), we observe similar signal and background probabilities for pA S2 $\sim 2500$ PE. Additionally, we have verified that the performance of a classifier trained with (pA S2, $\Delta t$) is significantly better than one trained with only (pA S2). This improvement is not observed for S1-related features; for example, the performance remains fairly similar whether using only (pA S1) or (pA S1, $\Delta t$).

The relevance of $\Delta t$ can also be seen in Fig.~\ref{fig:featureimportance} where we show the feature importance considering the gain metric for the XGBoost classifier. The most important features are both pA, the intensity of the signals, and $\Delta t$ in a close third place. The rest of the features seem not to be as relevant, however we noticed a significant performance drop if we train a classifier only using the 3 main features. We also need to include at least both pRMSW (dataset C3) to keep a similar performance.

Finally, in the second to last row of Table~\ref{tab:res} we show the results for the classifiers using dataset D composed of only two features: cS1 and cS2, two variables usually employed by the collaborations to perform their analyses. Recall that these quantities are related to the shape-related resumed features pA, the intensity of the S1 and S2 signals but corrected by geometric and response factors taking into account the particular characteristics of the experiment. From our results we can see that dataset D not only provides a simpler way to represent the data and ease the interpretation of the results, but also retains all the relevant information for the classification providing some of the highest AUC values. For that reasons cS1-cS2 is the best data representation. The only drawback is the determination of the correction factors, which implies a huge experimental effort. In that sense, dataset C3 composed of only five features that can be extracted directly from the raw signals, can be useful to cross-check the modeling of those factors.

\begin{figure}
  \centering
  \includegraphics[width=0.55\textwidth]{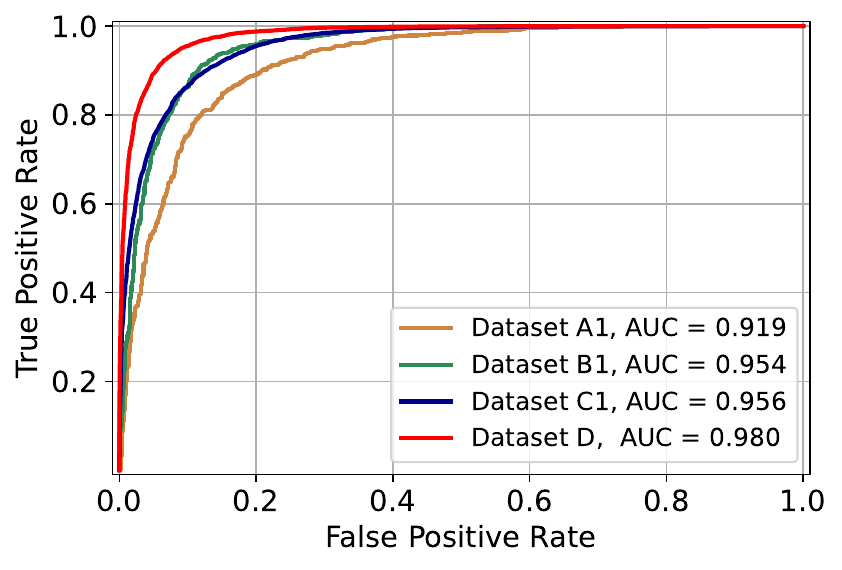}
\caption{ROC curves and AUC for several data representation examples. All curves consider XGBoost as classifier and a WIMP with $m_{dm}=500$ GeV as benchmark.}
\label{fig:rocs}
\end{figure}

Another important conclusion is that we have not found a significant difference in performance between the ML algorithms. 
Even more, the simple and very fast XGBoost method is usually on-par with more complex and harder to tune architectures. Additionally, it has the advantage of being more robust, in the sense that it is less sensitive to its hyperparameters, weight initialization, and yields a smaller uncertainty value.
In Fig.~\ref{fig:rocs} we show the ROC curves for some data representations and XGBoost, they correspond to ones with the lowest and highest AUC. The rest of the algorithms have similar ROC shapes, according to their AUC.

\subsection{Exclusion limits}

We computed the exclusion limits employing two different methods detailed in Appendix~\ref{Appen:stats} that combine ML and traditional statistical procedures. The process can be summarized as
\begin{itemize}
    \item Instead of dealing with the data representation original high-dimensional space, we work with the output of the binary classifiers described in previous sections, which is always one-dimensional.
    \item We employ two statistical methods, a binned and an unbinned analysis. In the first one, we divide the output of the ML in a histogram and estimate the expected number of signal and background events in each bin. In the second method, we estimate the signal and background probability density functions with a non-parametric algorithm. To do so, in both methods, we feed the already trained classifier with only signal events and then with only background events taken from an independent dataset, the test set.
    \item To set exclusion limits, we want to assess the compatibility of the signal-plus-background hypothesis when the true data has only background events. Therefore, we prepare a set of samples, or pseudo-experiments, each one containing the expected total number of background events, $B$, taken from the test dataset.
    \item We feed a ML classifier with a large number of pseudo-experiments, and compute for each one the test statistic $\tilde{q}_{\mu}$ of Eq.~\ref{eq:testexclusion1}. Notice that the pseudo-experiments do not depend on the DM characteristics, however the expected signal rate is incorporated into the test statistic formula. From the distribution of $\tilde{q}_{\mu}$ we obtain the significance, $Z$ (see Eq.~\ref{eq:Z1}).
    \item We fix $m_{dm}$~\footnote{In the test statistic formula, we use the expected signal rate that corresponds to the fixed DM mass value. Additionally, we employ a ML classifier that was trained to distinguish background vs signal events generated with the mentioned fixed DM mass. Later we will see that we can design a single ML algorithm that can be applied to a range of DM masses.} and compute the significance for different values of $\sigma^{SI}_{dm-\mathcal{N}}$. We set the exclusion limits at 90\% confidence level for the cross-section value with $Z=1.64$.
\end{itemize}

As mentioned in the previous section, the WIMP cross-section is not relevant to model the detector response, but is related to the expected number of events (see Appendix~\ref{Appen:signal_rates}). Therefore, to compute the limits in the cross-section vs mass plane we are going to employ the data generated with $\sigma^{SI-0}_{dm-\mathcal{N}}=10^{-45}$ cm$^2$ but re weighting the expected number of events for different values of cross-section as 
\begin{equation}
    S = S^0 \; \frac{\sigma^{SI}_{dm-\mathcal{N}} }{ \sigma^{SI-0}_{dm-\mathcal{N}} } ,
\end{equation}
where $S^0$ is the expected number of events for $\sigma^{SI-0}_{dm-\mathcal{N}}$.

It is crucial to highlight that since the statistic treatment to determine the exclusion limits is applied to the output of the ML binary classifier, the exclusion reach is determined by the performance of the mentioned classifier. This implies that among the 32 models presented in Table~\ref{tab:res}, we expect that the one with the highest AUC will yield the most stringiest limits. Additionally, we also expect that models with similar AUCs will yield similar constraints. As discussed in the previous section, the XGBoost method offers several advantages, including simplicity, speed and greater stability. Therefore, we present the exclusion limit analysis using only XGBoost classifiers. For a comparison of some examples using transformers see Appendix~\ref{Appen:exclusionlimits}, which illustrates that the choice of ML architecture results in negligible differences in the exclusion limit curves when the AUCs are similar.

\begin{figure}
  \centering
  \includegraphics[width=0.49\textwidth]{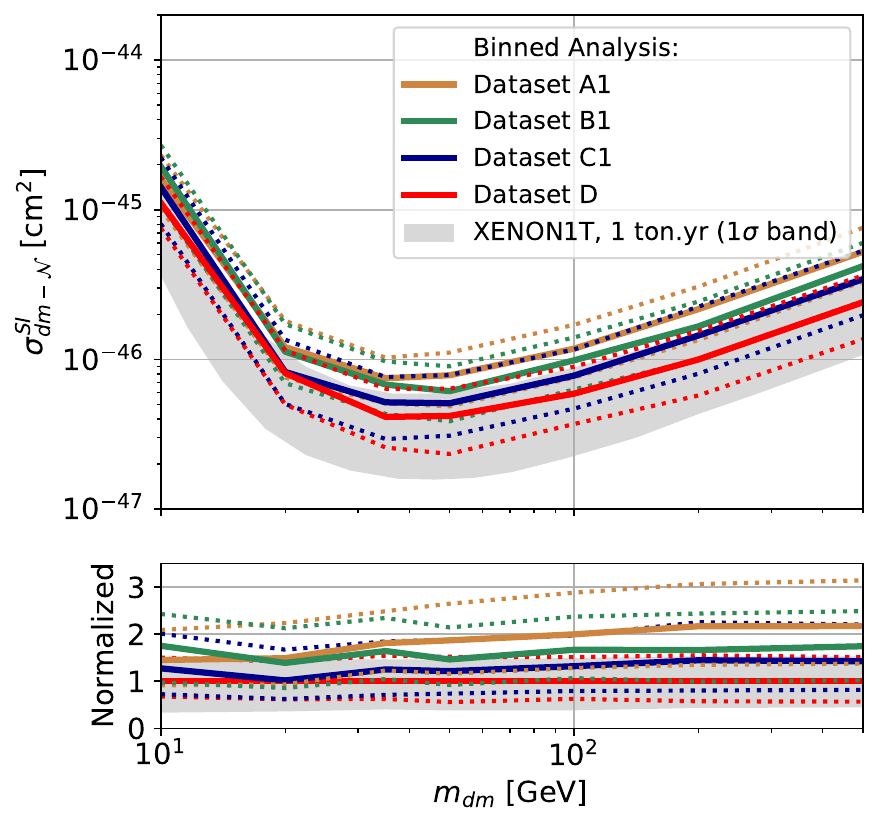}
  \includegraphics[width=0.49\textwidth]{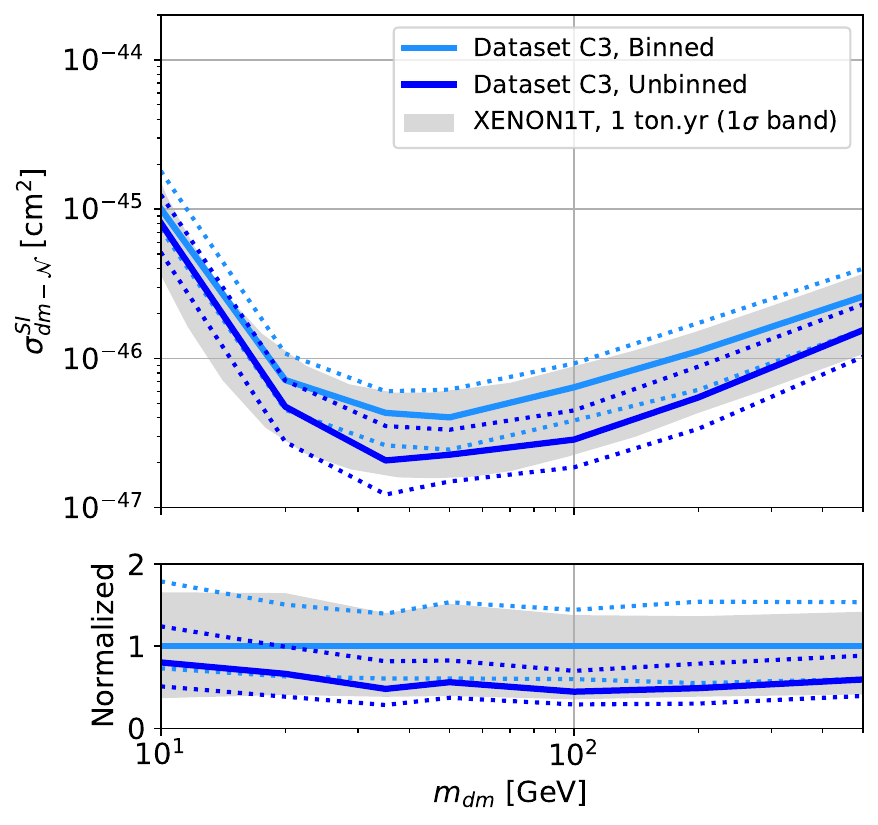}
\caption{Expected 90$\%$ C.L. upper limit on $\sigma^{SI}_{dm-\mathcal{N}}$ (solid curves) with 1$\sigma$ (dotted curves) sensitivity bands taking into account only statistical uncertainties. The result from XENON1T 1 ton yr~\cite{XENON:2018voc} is shown as a gray coloured area as reference. On the left panel we show examples with one of the highest AUC (dataset D), and the lowest AUC (dataset A1). On the right panel, we present a comparison between the binned and unbinned methods employed to compute the limits for the dataset C3. Every example considers XGBoost.}
\label{fig:limits}
\end{figure}

On the left panel of Fig.~\ref{fig:limits} we show the exclusion limits for some examples considering XGBoost as classifier. The solid lines represent the expected 90$\%$ C.L. upper limit on $\sigma^{SI}_{dm-\mathcal{N}}$, while the dotted curves the 1$\sigma$ sensitivity bands taking into account only statistical uncertainties. We also present as a gray band the expected limits set by XENON1T 1 ton yr~\cite{XENON:2018voc}. Although there are several differences between the analysis applied in this work and the one performed by the XENON collaboration (inclusion of all systematic uncertainties, data-driven analysis of the backgrounds, and most importantly the fact that the statistical treatment is carried out on the original cS1-cS2 plane instead of on the ML output of different representations), we can see that our results are compatible with XENON1T which validates the procedure and the simulator as an accurate tool to describe the behaviour of the detector. As anticipated, the model with one of the highest AUC values in Table~\ref{tab:res}, dataset D (AUC=0.980), provides the strongest constraints, while dataset A1 (AUC=0.919) the worst limits by a factor $\sim2$. We have checked that the rest of exclusion curves, considering other representations and ML algorithms, lie between those curves following the AUC order shown in Table~\ref{tab:res}.

Next we will focus on the dataset C3. This is one of the simplest representations with only 5 variables (4 main resumed features and $\Delta t$, the time between the pulses), that still provides one of the highest AUC and therefore a high exclusion limit reach. On the other hand, dataset D yields similar results with even less features (cS1 and cS2), and is the dataset usually employed by the experimental collaborations. However, it requires the determination of the correction factors to the original S1 and S2 signals, which implies a detailed detector characterization. This issue is absent with the dataset C3 case, as its features can be extracted directly from the raw signals. For completeness, in Appendix~\ref{Appen:exclusionlimits} we show the results using the well-known dataset D and compare them with the dataset C3. As expected, no significant differences are found.

Regarding the impact of employing a binned or an unbinned statistical treatment, on the right panel of Fig.~\ref{fig:limits} we show a comparison between the procedures using dataset C3. The unbinned method out-performs the binned one by a factor $\sim2$, and we have checked that this fact holds for every representation. Nonetheless, it should be noted that the unbinned method is significantly more computationally expensive compared to the binned case (it takes about one to two orders of magnitude longer to compute).

\begin{figure}
  \centering
  \includegraphics[width=0.49\textwidth]{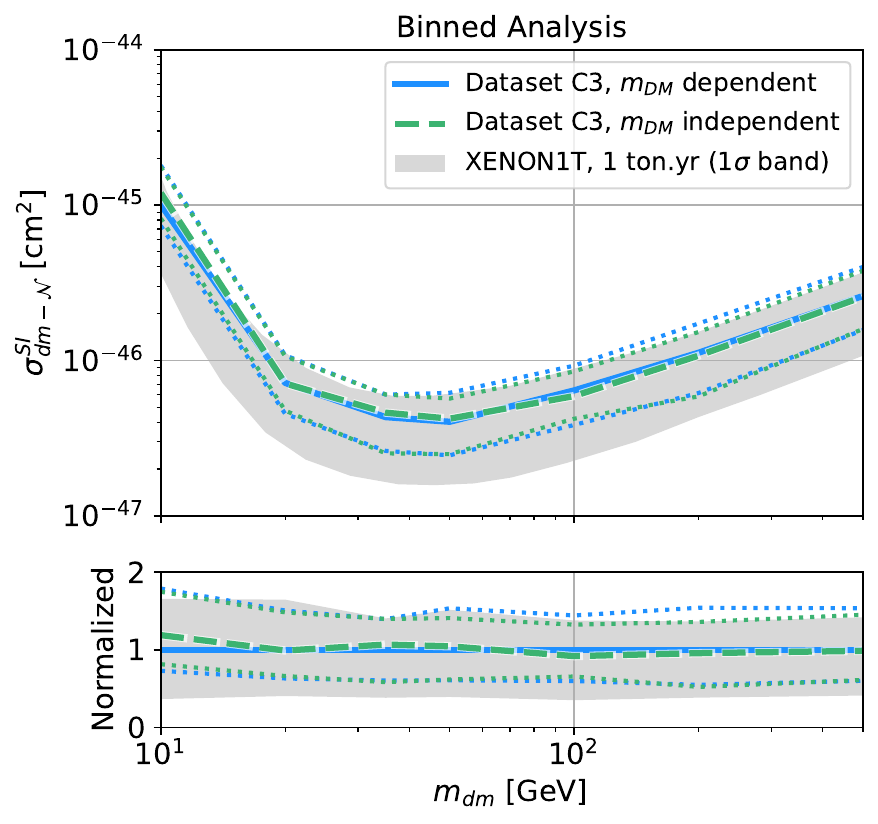}
  \includegraphics[width=0.49\textwidth]{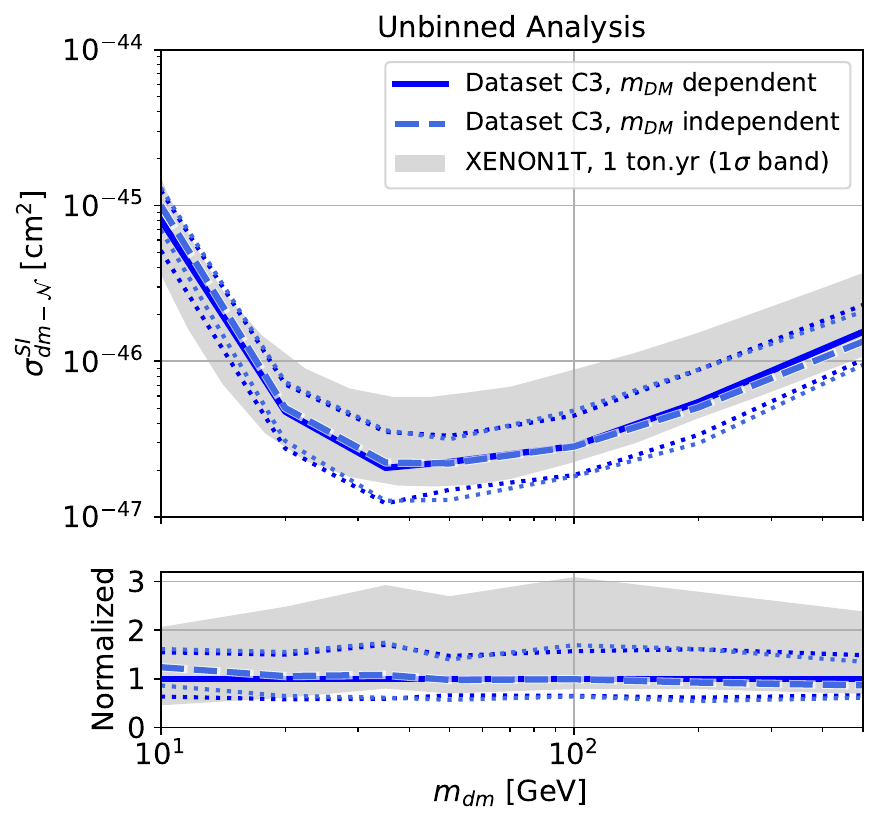}
\caption{Expected 90$\%$ C.L. upper limit on $\sigma^{SI}_{dm-\mathcal{N}}$ with 1$\sigma$ (dotted curves) sensitivity bands taking into account only statistical uncertainties. The result from XENON1T 1 ton yr~\cite{XENON:2018voc} is shown as a gray colored area as reference. The solid curves employ one specific ML classifier for each $m_{dm}$, while the dashed curves are obtained with a single ML classifier trained with a signal dataset combining all $m_{dm}$. The limits are computed with the binned (left panel) and unbinned (right panel) methods. Every example considers XGBoost and dataset C3.}
\label{fig:limits_general4}
\end{figure}

Finally, we would like to highlight that so far we have presented our procedure to an ideal scenario, i.e. computing the exclusion limits for a fixed DM mass using a specific ML classifier trained with a signal dataset generated with the same WIMP mass that we want to test. We call this approach $m_{dm}$ dependent. Since in a real experiment we do not know the true underlying DM model, we have trained a new ML classifier designed to distinguish between background events and a signal dataset generated with different WIMP masses, $m_{dm}=[10, 20, 35, 50, 100, 200, 500]$ GeV (equal number of events per mass). Then, we call this approach $m_{dm}$ independent.
Although a ML classifier trained with a set of masses may not be optimal compared with a ML classifier trained for a specific mass, the $m_{dm}$ independent approach would be the one useful in a real experiment since it is suited to be applied to a range of masses. In Fig.~\ref{fig:limits_general4} we show a comparison between both approaches, left panel for (the most efficient) binned analysis and right panel for (the best performing) unbinned one. We can see that for this particular physical problem, although not optimal, the $m_{dm}$ independent algorithm presents minimal differences with the $m_{dm}$ dependent method. This is related to both the excellent discrimination power of the classifiers that are able to learn the distribution of the backgrounds, and to the similarities of the signal when only the WIMP mass is changed.

\section{Conclusions}
\label{sec:conclusions}

In this study, we performed an extensive comparison between deep learning models, such as state-of-the-art transformers, traditional methods like Multilayer Perceptrons and Convolutional Neural Networks, and boosted decision models like XGBoost to improve the detection capabilities of liquid xenon time projection chamber experiments. We evaluated different data representations and found that simplified feature representations, especially dataset D (corrected S1 and S2 signals), and dataset C3 (4 main resumed features $+$ $\Delta t$, the time difference between S1 and S2), are effective in obtaining important information for classification tasks. However, the latter representation avoids the need to determine the correction factors obtained through detailed detector characterization, and can be used as a simple cross-check.

Our results show that while complex transformer models offer promising performance, simpler and significantly faster models such as XGBoost can achieve comparable results when optimal data representations are used. This emphasizes the importance of feature engineering and data representation in machine learning applications for DM detection.

We also derived exclusion bounds in the cross-section versus DM mass parameter space ($\sigma^{SI}_{dm-\mathcal{N}}$ vs. $m_{dm}$), revealing minimal differences between the performance of XGBoost and deep learning models. This suggests that simpler models, when optimized accordingly, can be as effective as more complex models for this particular application.


\begin{acknowledgments}

The authors would like to thank Susana Cebri\'an Guajardo and El\'ias L\'opez Asamar for useful discussions and comments on the manuscript.
The author(s) gratefully acknowledges the computer resources at Artemisa, funded by the European Union ERDF and Comunitat Valenciana as well as the technical support provided by the Instituto de Física Corpuscular, IFIC (CSIC-UV).
The work of D.L. was supported by the Argentinian CONICET, and he also acknowledges the support through {PICT~2020-02181}.  
The work of A.P. was supported by the Spanish Research Agency (Agencia Estatal de Investigaci\'on) through the Grant IFT Centro de Excelencia Severo Ochoa No CEX2020-001007-S and PID2021-125331NB-I00, funded by MCIN/AEI/10.13039/501100011033. AP also acknowledges support from the Comunidad Aut\'onoma de Madrid and Universidad Aut\'onoma de Madrid under grant SI2/PBG/2020-00005. R. RdA is supported by PID2020-113644GB-I00 from the Spanish Ministerio de Ciencia e Innovación and by the PROMETEO/2022/69 from the Spanish GVA.

\end{acknowledgments}


\appendix

\section{Signal rates}
\label{Appen:signal_rates}

To compute the expected event rate from the elastic scattering of a DM particle, we need the flux of incoming particles and the differential cross-section with the target. The dark matter flux can be expressed as
\begin{equation}
    \frac{d\Phi_{dm}}{d\bm{v}} = \frac{\rho_{dm}}{m_{dm}} \, v \, f(\bm{v},t) ,
\end{equation}
where $\rho_{dm}$ is the local DM density, $\bm{v}$ the relative velocity between the DM particle and the target, $f(\bm{v},t)$ the DM velocity distribution in the detector frame normalized to unity for which we assume the Standard Halo Model (SHM) a Maxwell-Boltzmann velocity distribution, with velocity dispersion 238 km/s~\cite{Koposov_2010} and truncated at the escape velocity from the Milky Way, $v_{esc}=533$ km/s~\cite{Piffl:2013mla}.

Considering the spin-independent scattering case and equal couplings of DM to protons and neutrons, the differential cross-section for DM-target scattering takes the form
\begin{equation}
\begin{split}
    \frac{d\sigma_{dm-T}}{dE_R} & = \frac{\sigma_{dm-T}}{E_R^{max}} \\
    & =  \frac{\sigma^{SI}_{dm-\mathcal{N}}}{E_R^{max}} \, \frac{\mu_{dm-T}^2}{\mu_{dm-\mathcal{N}}^2} \, A_T^2 \, F^2(E_R) \\
    & = \frac{ m_T \, \sigma^{SI}_{dm-\mathcal{N}} \, A_T^2 }{2 \, \mu_{dm-\mathcal{N}}^2 \, v^2} \, F^2(Q) ,
\end{split}
\end{equation}
where $\sigma^{SI}_{dm-\mathcal{N}}$ is the spin-independent DM-nucleon scattering cross-section at zero-momentum,
$A_T$ is the atomic mass number of the target nucleus, $\mu_{dm-T}$ ($\mu_{dm-\mathcal{N}}$) is the DM-target (DM-nucleon) reduced mass, $F(Q)$ is the nuclear form factor~\cite{PhysRev.104.1466} which accounts for the finite nuclear size and loss of coherence related to the momentum transfer $Q = \sqrt{2 m_T E_R}$. From kinematics, the maximum transferred energy in a single elastic scattering by a DM particle of kinetic energy $T_{dm}$ is
\begin{equation}
\begin{split}
    E_R^{max} & = \frac{2 \, m_T T_{dm} \, (T_{dm} + 2 \, m_{dm})}{2 \, m_T \, T_{dm} + (m_{dm} + m_T)^2} \\
    & \simeq \frac{4 \, m_T \, m_{dm}}{(m_{dm} + m_T)^2} \, T_{dm} \\
    & = \frac{2 \, \mu_{dm-T}^2 \, v^2}{m_T} ,
\end{split}
\end{equation}
where $m_T$ is the mass of the nucleus target, and we used a non-relativistic WIMP,
$v \sim 10^{-3}$, $T_{dm} = \frac{1}{2} m_{dm} v^2 \ll m_{dm}, m_T$.

Then, the differential rate of nuclear recoils with energy $E_R$ is given by
\begin{equation}
\begin{split}
    \frac{dR}{dE_R} & = \frac{\epsilon_T}{m_T} \, \int_{v_{min}}^{v_{esc}} d\bm{v} \, \frac{d\Phi_{dm}}{d\bm{v}} \, \frac{d\sigma_{dm-T}}{dE_R}  \\
    & = \epsilon_T \, \frac{\sigma^{SI}_{dm-\mathcal{N}} \, A_T^2 \, F^2(E_R)}{2 \, \mu_{dm-\mathcal{N}}^2}  \frac{\rho_{dm}}{m_{dm}} \, \int_{v_{min}}^{v_{esc}} d\bm{v} \, \frac{f(\bm{v},t)}{v} ,
\end{split}
\label{eq:diff_rate}
\end{equation}
where $\epsilon_T$ is the total exposure, given by the product of the detector mass and the run-time, hence $\frac{\epsilon_T}{m_T}$ is the number of targets times the run-time. The integration over the DM velocity takes into account the minimum DM speed needed to induce a nuclear recoil of energy $E_R$, $v_{min}(E_R) = \sqrt{\frac{m_T \, E_R}{2 \, \mu_{dm-T}^2}}$.

Finally, the expected event rate is found as
\begin{equation}
    N = \int_{E_R^{min}}^{E_R^{max}} dE_R \, \eta(E_R) \, \frac{dR}{dE_R} ,
\end{equation}
with $\eta(E_R)$ the energy-dependent detector efficiency, and $E_R^{min}$ ($E_R^{max}$) the minimum (maximum) recoil energies that define the region of interest of the experiment, typically from a few keV to almost 100 keV for liquid xenon TPCs.

\section{Machine Learning training details}
\label{Appen:MLtraining}

In this appendix we provide more details on the architecture and training of the ML classifiers. Since we studied 32 independently tuned data-algorithm combinations (not including several tests and combinations not shown in Table~\ref{tab:res}) in this section we present the details of some representative cases. For further information about the architectures, algorithm and dataset, the code used in this work is available at~\cite{DDcode}.

\begin{figure}
  \centering
  \includegraphics[width=0.49\textwidth]{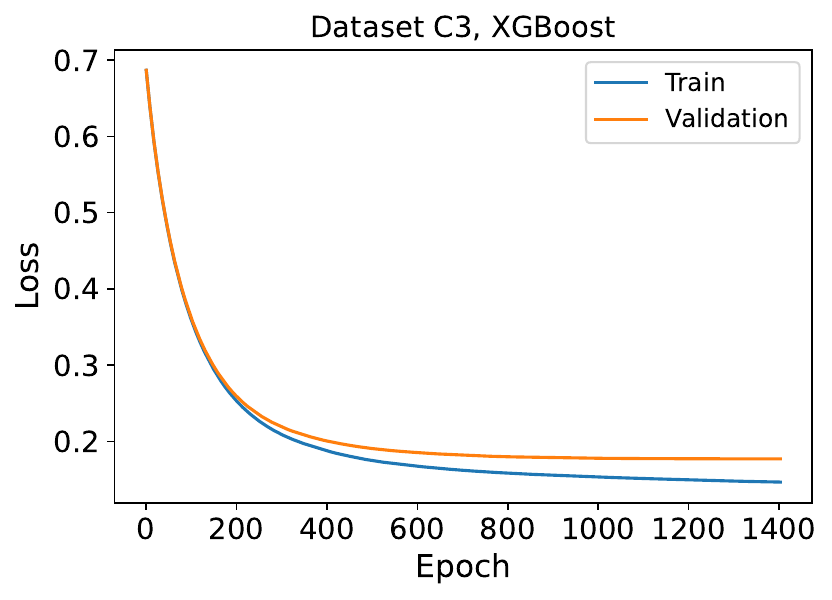}
  \includegraphics[width=0.49\textwidth]{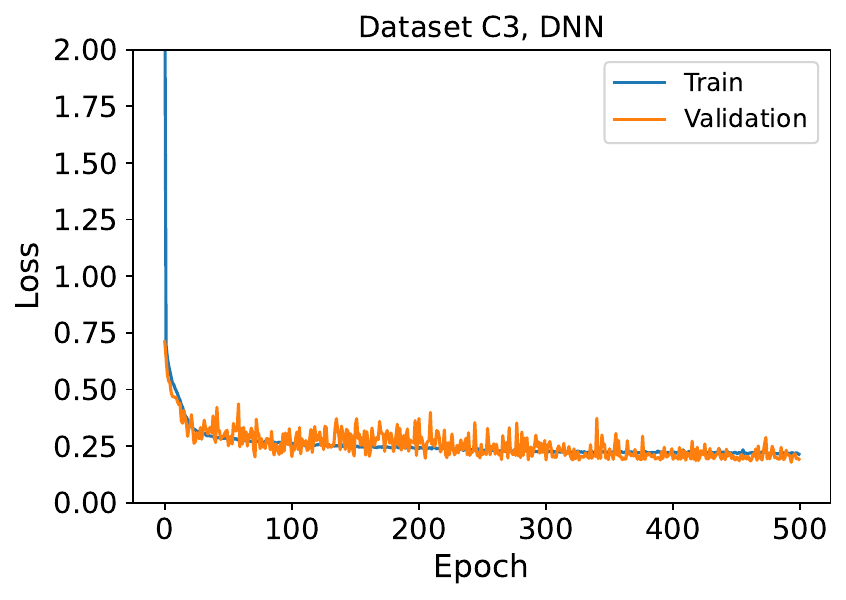}
  \includegraphics[width=0.49\textwidth]{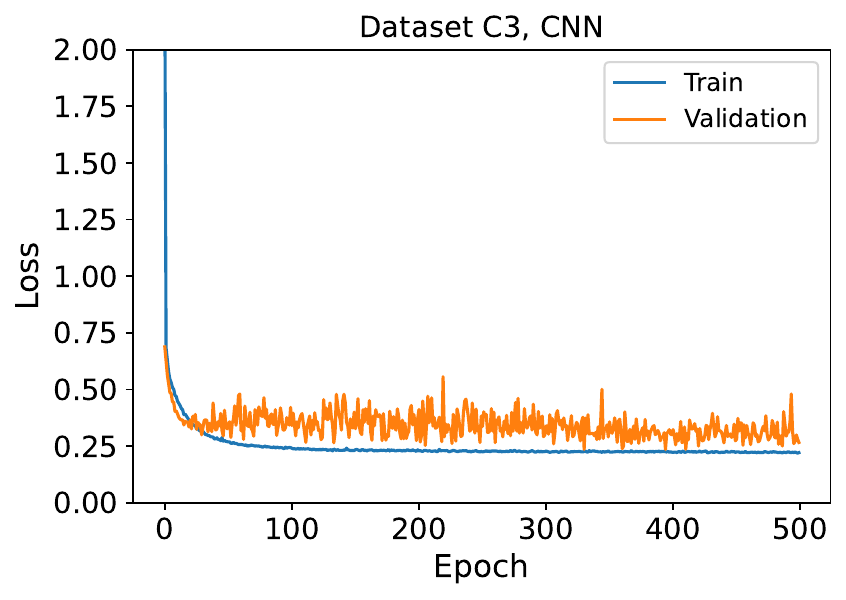}
  \includegraphics[width=0.49\textwidth]{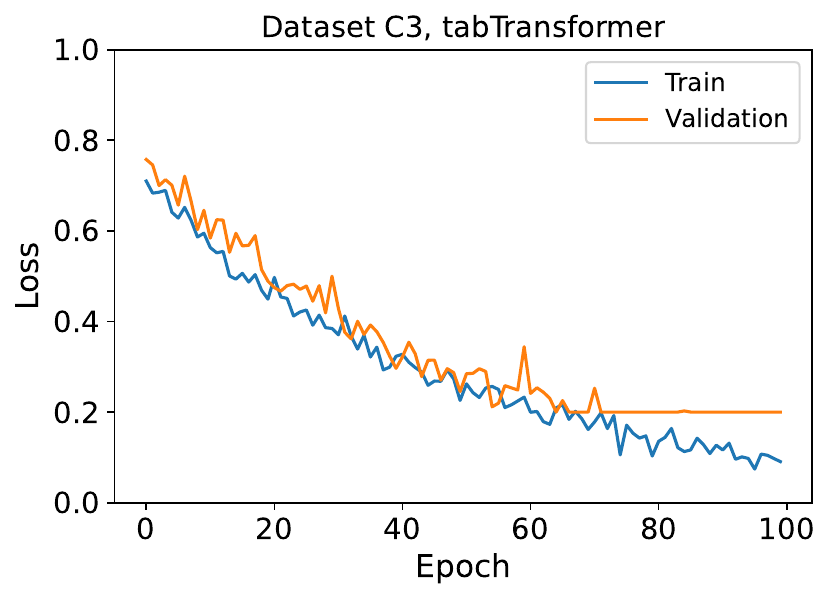}
\caption{Training and validation loss as a function of the training epoch for the data representation dataset C3. Each panel corresponds to one of the ML algorithms employed in this work.}
\label{fig:loss}
\end{figure}

In Fig.~\ref{fig:loss} we show the train and validation loss as a function of the epoch during training, for the data representation dataset C3. To avoid overfitting an early stopping of 50 iterations was used in every ML algorithm to end the training procedure if the validation performance does not improve for the last given amount of epochs. Additionally, the command \texttt{restore$\_$best$\_$weights} was set to \texttt{True} to save and use the weights from the best epoch during training (the one with the best performance), instead of using the weights from the last epoch that might be sub-optimal if training continued past the ideal point.

The XGBoost architecture has 12 hyperparameters, that we have tuned using a grid search. For the shown example in Fig.~\ref{fig:loss}, the most relevant ones for our classification task are a maximum depth of a tree equal to 5, learning rate of 0.01, 2500 estimators, and a L2 regularization term on weights of 0.01, although we found that the default hyperparameter values provide near optimal results. As learning objective we choose \texttt{binary:logistic} and \texttt{logloss} as evaluation metric.

For the DNN example, we used 6 hidden layers with 128, 64, 32, 16, 8, and 4 units per layer, respectively, all with Leaky-ReLU as activation function with alpha of 0.05. These parameters, as well as the tuning of the activation function, have a significant impact in the network performance. Additionally, we set a dropout value of 0.2 after each layer, used a sigmoid function as activation in the final layer, defined \texttt{binary$\_$crossentropy} as loss function for the training, using \texttt{Adam} as optimizer with its default values, and a batch size of 128. The number of training epoch was set high enough such that the early stopping algorithm finishes the fitting procedure. This results is a  classifier with a total of 11809 trainable parameters.

The optimal CNN found for this example shares some hyperparameters with the mentioned DNN: the same optimizer, batch size, dropout values, and the same Leaky-ReLU activation function. We used 2 1D convolutional layers (since in this case we are using tabulated data) with 32 filters, and a kernel size of 3 and  2 for the first and second convolutional layer, respectively, a stride equal to 1, no padding, and an L2 kernel regularizer with value 0.001. After the convolutional layers we set a 1D max-pooling layer of size equal to 2, a dropout of 0.2, a flatten layer, and a fully connected layer of 32 units and Leaky-ReLU as activation function with alpha of 0.05, and dropout of 0.2. Finally, a sigmoid function as activation in the final layer was employed. Then, a network with a total of 3297 trainable parameters was generated.

The tabTransformer hyperparameters are set as follows: the embedding dimension (\texttt{input$\_$embed$\_$dim}) is fixed at 128, which defines the size of the input embeddings for the features, providing a balance between model complexity and representational capacity. The number of attention heads (\texttt{num$\_$heads}) is set to 16, enabling the model to attend to different parts of the input representation simultaneously. Lastly, with \texttt{num$\_$attn$\_$blocks}=12, the model uses 12 stacked attention layers to capture hierarchical relationships between features while maintaining reasonable model depth.

\begin{figure}
  \centering
  \includegraphics[width=0.75\textwidth]{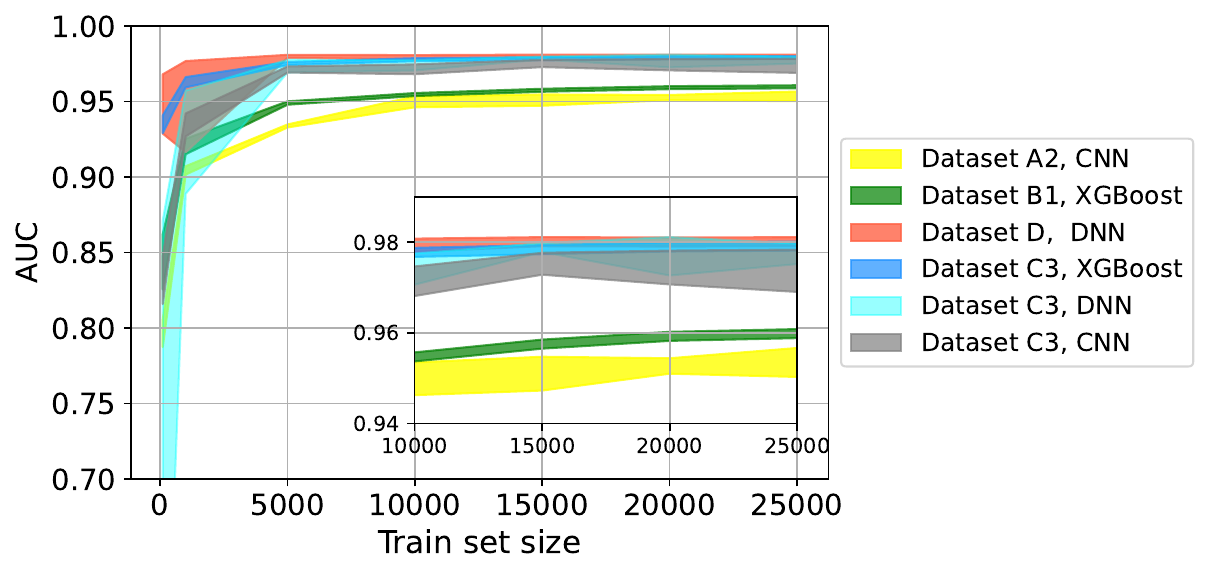}
\caption{AUC as a function of the train dataset size for several data representation and ML algorithms examples. The train set size includes background and signal events. The width represents 1$\sigma$ error from resampling.}
\label{fig:aucs}
\end{figure}

To determine the dataset size needed, we performed a systematic study comparing the results using datasets of different length. For each ML classifier shown in Table~\ref{tab:res}, we computed the AUC 5 times per dataset length considered. We report the mean and one standard deviation. In Fig.~\ref{fig:aucs} we show the AUC as a function of the train dataset size (which includes both background and signal events) for a few representative examples, noting that the rest of the ML classifiers follow the same tendency. As we can see, the AUC results are stable for datasets with $\gtrsim 10k - 15k$ events per class.

\section{Statistical treatment}
\label{Appen:stats}

\subsection*{Exclusion Limits}

We summarize the main features of the statistical treatment employed in this work based on the frequentist significance test using a likelihood ratio as a test statistic. It allows obtaining exclusion limits, comparing a null hypothesis (the signal-plus-background one) against an alternative one (the background-only one). For further details, see Ref.~\cite{Cowan:2010js}. 

We start by defining the statistic
\begin{equation}
    t_{\mu} = -2 \, \ln \frac{\mathcal{L}(\mu,\mathbf{x})}{\mathcal{L}(\hat{\mu},\mathbf{x})}
\end{equation}
where $\mathcal{L}(\mu,\mathbf{x})$ is the likelihood function describing a set of independent measurements with arbitrarily high-dimensional set of observables $\mathbf{x}$, the parameter $\mu$ determines the strength of the signal process, $\mu=0$ corresponds to the background-only hypothesis and $\mu=1$ signal-plus-background hypothesis, and the value $\hat{\mu}$ maximizes the likelihood.

The likelihood ratio close to 1 implies good agreement between the data and the hypothesized value of $\mu$, otherwise the ratio is close to 0. Then $0 \leq t_{\mu} \leq \inf$, with higher values of $t_{\mu}$ corresponding to a greater incompatibility between data and $\mu$. 

In this work, we assume that the presence of the signal can only increase the mean event rate beyond what is expected from background, $\mu \geq 0$. Therefore, in case of data such that $\hat{\mu} < 0$,  the best possible physical value of $\mu$ giving the best agreement with the data is $\mu=0$. On the other hand, for exclusion limits, if $\hat{\mu} > \mu$ one would not regard that data to be incompatible with the considered $\mu$ and then is not taken as part of the rejection region by fixing the test to 0 (recall that $t_{\mu}=0$ indicates agreement between data and the hypothesis). For those reasons, it is convenient to define the following test statistic
\begin{equation}
\tilde{q}_{\mu} =   \begin{cases}
    0       & \quad \text{if } \hat{\mu} > \mu\,,\\
    -2 \, \ln \frac{\mathcal{L}(\mu,\mathbf{x})}{\mathcal{L}(\hat{\mu},\mathbf{x})}       & \quad \text{if } 0 \leq \hat{\mu} \leq \mu \,,\\
    -2 \, \ln \frac{\mathcal{L}(\mu,\mathbf{x})}{\mathcal{L}(0,\mathbf{x})}  & \quad \text{if } \hat{\mu} < 0\,.
  \end{cases}
  \label{eq:testexclusion1}
\end{equation}

We can estimate numerically the test statistic with a finite dataset $\mathbb{M}$. The probability distribution of $\tilde{q}_{\mu}$ is going to be conditioned to the true signal strength $\mu'$ that defines the mentioned dataset. Since to set exclusion limits we want to assess the compatibility of signal-plus-background hypothesis when the true data has only background events, we construct a dataset $\mathbb{M}$ containing only background events, i.e. $\mu'=0$. In practice, we generate a large number of pseudo-experiments~\footnote{For exclusion, a pseudo-experiment is a set of background-only events such that its number of events is taken from a Poisson distribution with mean equal to the expected number of background events for the considered experiment.} taken from the dataset $\mathbb{M}$, and compute for each one the value of $\tilde{q}_{\mu}$ of Eq.~\ref{eq:testexclusion1}. With that procedure, we can obtain numerically the distribution of $[\tilde{q}_{\mu}|\mu'=0]$ and the median expected exclusion significance $\text{med }[Z_{\mu}| \mu'=0]$ is defined as
\begin{equation}
    Z = \text{med }[Z_{\mu}|\mu'=0] = \sqrt{\text{med }[\tilde{q}_{\mu}|\mu'=0]}\, , 
    \label{eq:Z1}
\end{equation}
and its statistical uncertainty can be estimated as
\begin{equation}
    \sigma_Z = \frac{1}{2 \sqrt{\text{med }[\tilde{q}_{\mu}|\mu'=0]}} \, \sigma_{[\tilde{q}_{\mu}|\mu'=0]} \, .
\end{equation}
To impose exclusion limits at 90\% confidence level we consider $Z=1.64$ according to the one-sided definition.

\subsection*{Machine Learning in statistical tests}

To deal with data of arbitrarily high dimensions $\mathbf{x}$, we employ a ML classifier and work with its output, the classification score $o(\mathbf{x}) \in [0,1]$ that maximizes the binary cross-entropy. 

If we divide the ML output into $D$ bins, the likelihood function can be represented as the product of Poisson distributions
\begin{equation}
    \mathcal{L}(\mu,o(\mathbf{x})) = \prod_{d=1}^{D}\text{Poiss}\big(N_{d}|\mu S_{d} + B_{d}\big) \, .
\label{eq:binned-likelihood}
\end{equation}
where $S_{d}$ ($B_{d}$) is the expected number of signal (background) events, and $N_{d}$ the measured number of events, in the bin $d$. This approach is called Binned Likelihood (BL) method. We would like to highlight that there is a well-known problem with this approach, called the curse of dimensionality, that states that to reliably populate the bins one needs to increase the number of data points exponentially with the dimension of the observable. However, if the observable is a ML output $o(\mathbf{x})$ the dimension is always 1, regardless of the original space $\mathbf{x}$, which allows the application of the BL method in significantly more scenarios.

The test statistic in Eq.~\eqref{eq:testexclusion1} for the BL approach is
\begin{equation}
\tilde{q}_{\mu} =   \begin{cases}
0 & \rm{if} \, \, \hat{\mu} > \mu \\
2 \sum_{d=1}^{D} (\mu - \hat{\mu})S_d - N_d \ln \left( \frac{\mu S_d + B_d}{ \hat{\mu} S_d + B_d} \right) &  \rm{if} \, \, 0 \leq \hat{\mu} \leq \mu \\
2 \sum_{d=1}^{D} \mu S_d - N_d \ln \left( 1+ \frac{\mu S_d}{B_d} \right)  &  \rm{if} \, \,  \hat{\mu} < 0;  
  \end{cases}
\label{eq:testexclusionBL}
\end{equation}
and $\hat{\mu}$ has to be extracted from the data maximizing Eq.~\ref{eq:binned-likelihood} 
\begin{equation}
    \sum_{d=1}^{D}\frac{N_d S_d}{\hat{\mu}S_d +B_d} - S_d = 0 \, .
\label{eq:muhatBL}
\end{equation}

We can also describe our data with an unbinned likelihood
\begin{equation}
    \mathcal{L}(\mu,o(\mathbf{x})) = \text{Poiss}\big(N|\mu S + B\big)\,\prod_{i=1}^{N}p(o(\mathbf{x}_i)|\mu,s,b) \, ,
\label{eq:stat_model} 
\end{equation}
where $S$ ($B$) is the total expected signal (background) yield, $N$ the total number of independent measurements, and $p(o(\mathbf{x}_i)|\mu,s,b)$ is the probability density for a single measurement $o(\mathbf{x}_i)$ containing the event-by-event information as a mixture of signal and background densities
\begin{equation}
    p(o(\mathbf{x}_i)|\mu,s,b) = \frac{B}{\mu S + B}\,p_{b}(o(\mathbf{x}_i))+\frac{\mu S}{\mu S + B}\,p_{s}(o(\mathbf{x}_i)) \, ,
\label{eq:prob_single_measurement}
\end{equation}
where $p_{s}(o(\mathbf{x}))=p(o(\mathbf{x})|s)$ and $p_{b}(o(\mathbf{x}))=p(o(\mathbf{x})|b)$ are the signal and background probability density functions (PDFs) for a single measurement $o(\mathbf{x})$, respectively, with $\frac{\mu S}{\mu S + B}$ and $\frac{B}{\mu S + B}$ the probabilities of an event being sampled from the corresponding probability distributions. 

Since $p_{s,b}(o(x))$ are the distributions of $o(x)$ for signal and background, we can obtain them by evaluating the classifier on a set of pure signal or background events, respectively. To extract them we used a non-parametric method called Kernel Density Estimation (KDE)~\cite{RosenblattKDE,ParzenKDE}, that has the advantage of not
assuming any functional form for the distributions, through its \texttt{scikit-learn} implementation \cite{scikit-learn}. The KDE method also suffers from the curse of dimensionality, however we are applying the algorithm to the one-dimensional output of the machine learning classifier, avoiding this issue.

Finally, the test statistic in Eq.~\eqref{eq:testexclusion1} becomes
\begin{equation}
\tilde{q}_{\mu} =   \begin{cases}
0 & \rm{if} \, \, \hat{\mu} > \mu \\
2(\mu-\hat{\mu}) S - 2 \sum_{i=1}^{N} \ln \left( \frac{\mu S p_{s}(o(\mathbf{x}_{i})) + B p_{b}(o(\mathbf{x}_{i}))}{\hat{\mu} S p_{s}(o(\mathbf{x}_{i})) + B p_{b}(o(\mathbf{x}_{i}))}\right)  &  \rm{if} \, \, 0 \leq \hat{\mu} \leq \mu \\
2\mu S - 2 \sum_{i=1}^{N} \ln \left( 1 + \frac{\mu S p_{s}(o(\mathbf{x}_{i}))}{B p_{b}(o(\mathbf{x}_{i}))}\right)  &  \rm{if} \, \,  \hat{\mu} < 0;  
  \end{cases}
\label{eq:testexclusion2}
\end{equation}
and $\hat{\mu}$ that maximize Eq.~\ref{eq:stat_model} can be expressed as
\begin{equation}
    \sum_{i=1}^{N}\frac{p_{s}(o(\mathbf{x}_{i}))}{\hat{\mu}S\, p_{s}(o(\mathbf{x}_{i})) +B\, p_{b}(o(\mathbf{x}_{i}))} = 1 \, .
\label{eq:muhat}
\end{equation}

For more details and examples of implementation of the ML approach in statistical tests, we refer the reader to Ref.~\cite{Arganda:2022qzy,Arganda:2022mrd,Arganda:2022zbs,Arganda:2023qni}.

\section{Comparison of exclusion limits for several methods}
\label{Appen:exclusionlimits}

\begin{figure}
\centering
  \includegraphics[width=0.49\textwidth]{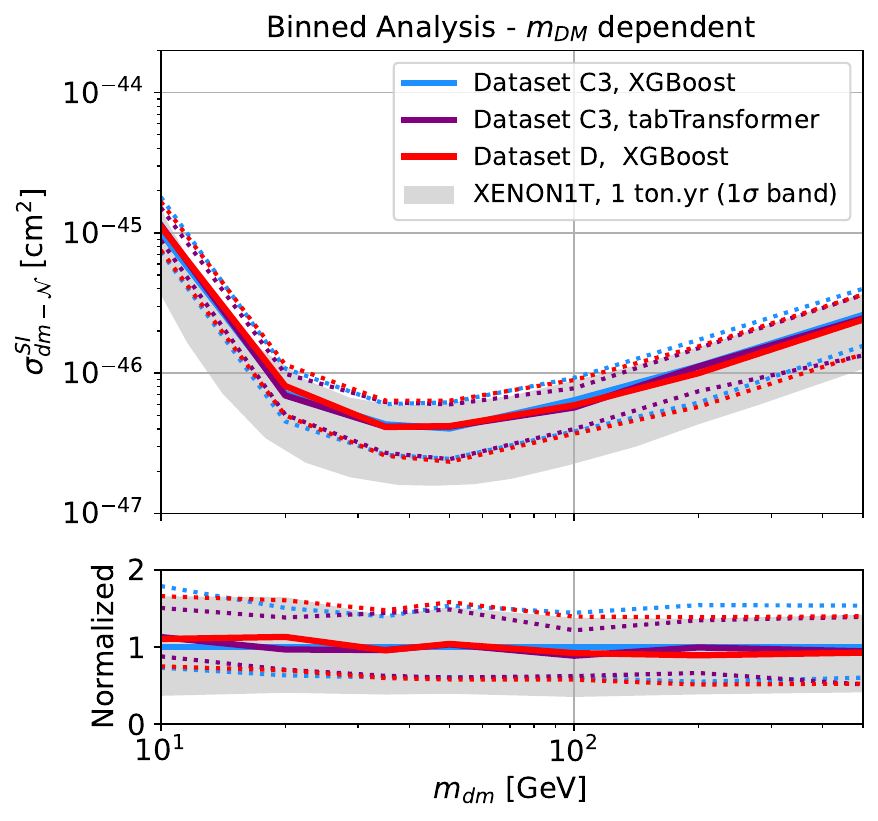}
  \includegraphics[width=0.49\textwidth]{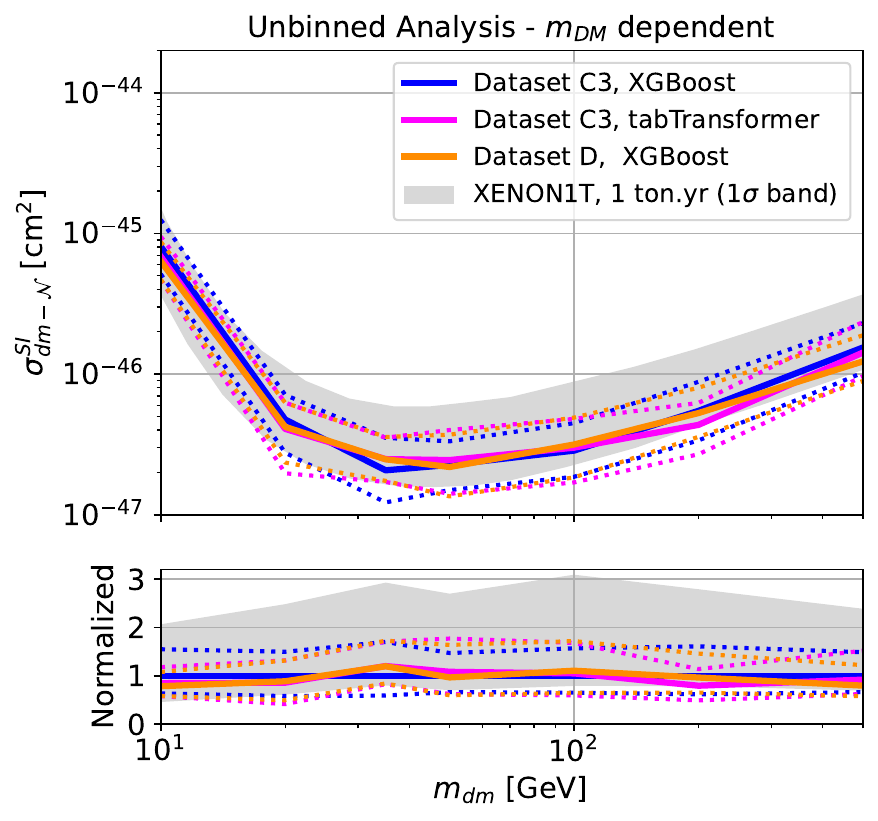}\\
  \vspace{0.5cm}
  \includegraphics[width=0.49\textwidth]{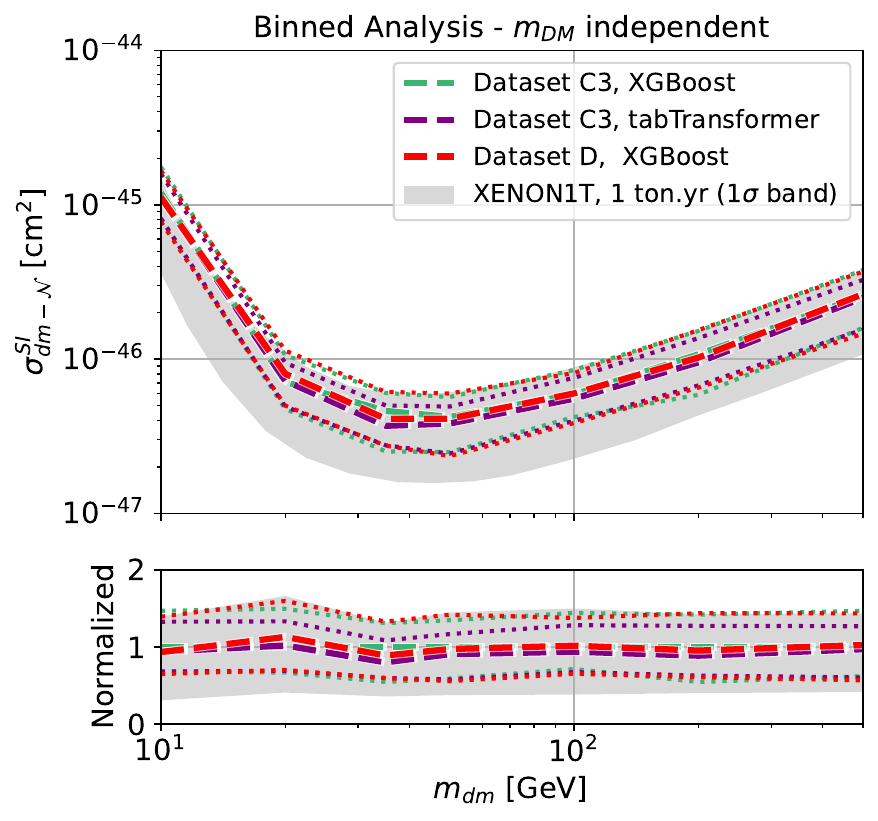}
  \includegraphics[width=0.49\textwidth]{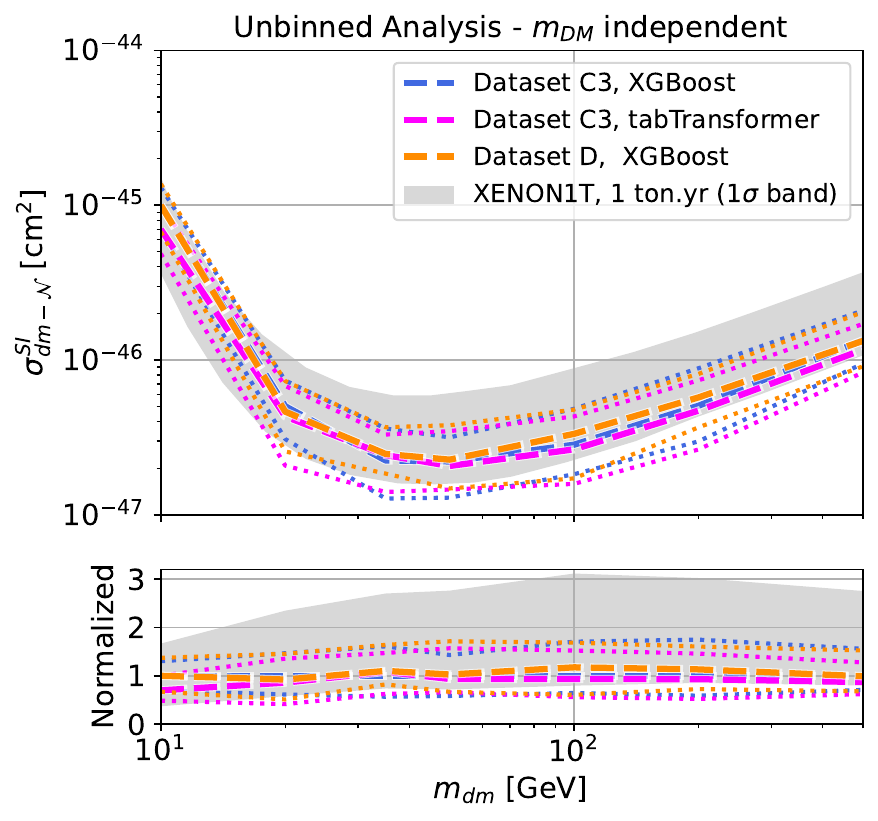}
\caption{Expected 90$\%$ C.L. upper limit on $\sigma^{SI}_{dm-\mathcal{N}}$ with 1$\sigma$ (dotted curves) sensitivity bands taking into account only statistical uncertainties. The result from XENON1T 1 ton yr~\cite{XENON:2018voc} is shown as a gray colored area as reference. The limits are computed with the binned (first column) and unbinned (second column) methods. The first row is obtained with one specific ML classifier for each $m_{dm}$ ($m_{dm}$ dependent method), while the second row a single ML classifier trained with a signal dataset combining all $m_{dm}$ ($m_{dm}$ independent method). In every panel we compare the results using the data representation dataset C3 with XGBoost and tabTransformers, and dataset D with XGBoost.}
\label{fig:limits_compare}
\end{figure}

In this appendix we compare the results for different data representation, ML algorithm and statistical treatment combinations. 
The exclusion limits shown on the first column of Fig.~\ref{fig:limits_compare} are computed using the binned method, while the second column presents the results with the unbinned procedure. The first row corresponds to the $m_{dm}$ dependent approach, where one specific ML classifier for each $m_{dm}$ is trained, while the second row shows the $m_{dm}$ independent strategy, using a single ML classifier and a single signal dataset that combines all the $m_{dm}$ sets.

In particular, we compare the results using the data representation dataset C3 with XGBoost and tabTransformers, and dataset D with XGBoost. 
The first data representation is the simplest one since it is composed of 5 features that require only basic operations to be extracted from the detector response. Additionally this representation is among the best performing ones. For this case, we compare the simplest and fastest XGBoost algorithm against an state-of-the-art transformer. On the other hand, the second data representation is constructed with the well-known cS1, cS2 variables employed in the experimental collaborations analyses. It is composed of only 2 features, but requires a thorough detector characterization to determine the correction factors to the original S1 and S2 signals. As we can see from the figure, the simplest data-ML architecture combination provides the same exclusion power as the more complex ones.


\bibliographystyle{utphys}
\bibliography{DM-ML-v3.bbl}

\end{document}